\documentclass[manuscript,screen]{acmart}

\usepackage{url} 
\usepackage[utf8]{inputenc} 
\usepackage{booktabs} 
\usepackage{graphicx}
\usepackage{tikz}
\usepackage{pgfplotstable}
\usepackage{svg}

\usepackage{quoting, lipsum}

\usepackage{wrapfig}

\usetikzlibrary{patterns}
\pgfplotsset{compat=1.17}

\usepackage{amsmath}

\usepackage{makecell}
\usepackage{rotating, multirow}

\usepackage{listings}
\usepackage{color}

\lstdefinelanguage{rock}{
	keywords={SUM,IFF,GROUP,BY},
    morekeywords = {for,each,FOR,ALL,EACH,WHERE,AFTER,PRECEDES,COUNT,ASSERT,IN,SELECT,AND,LATEST},
	keywordstyle=\color{blue},
	moredelim=[is][\color{dkgreen}\textbf]{|}{|},
	basicstyle=\footnotesize\ttfamily\linespread{0.1},
	morestring=*[d]{"},
	stringstyle=\color{redstrings},
	escapeinside={(*}{*)},
	columns=fullflexible,
}

\usepackage{csquotes}

\usepackage{float}

\usepackage{balance}

\usepackage{tabularx}
\usepackage{boxedminipage}
\newcolumntype{Y}{>{\centering\arraybackslash}X}
\newcolumntype{M}{>{\centering\arraybackslash}m}
\newcolumntype{S}{>{\centering\arraybackslash}p{0.085\textwidth}}
\newcolumntype{C}{>{\centering\arraybackslash}c}

\usepackage{multirow}
\usepackage{url}
\usepackage{mdframed}

\usepackage{amsmath}

\newcommand{\wk}[1]{{\textcolor{red}{~[~\textbf{WK}: \textit{#1} ]}}}

\AtBeginDocument{%
  }

\setcopyright{acmlicensed}
\copyrightyear{2024}
\acmYear{2024}
\acmDOI{XXXXXXX.XXXXXXX}

\acmConference[Conference acronym 'XX]{Make sure to enter the correct conference title from your rights confirmation emai}{June 03--05, 2018}{Woodstock, NY}
\acmISBN{978-1-4503-XXXX-X/18/06}

\begin{document}

\title{An Empirical Study on Challenges of Event Management in Microservice Architectures}


\author{Rodrigo Laigner}
\orcid{0000-0003-2771-7477}
\affiliation{%
	\institution{University of Copenhagen}
	\city{Copenhagen}
	\country{Denmark}
}
\email{rnl@di.ku.dk}

\author{Ana Carolina Almeida}
\authornote{Work done while employed as a postdoc at the University of Copenhagen.}
\orcid{0000-0003-0936-1542}
\affiliation{%
	\institution{State University of Rio de Janeiro}
	\city{Rio de Janeiro}
	\country{Brazil}
}
\email{ana.almeida@ime.uerj.br}

\author{Wesley K. G. Assunção}
\orcid{0000-0002-7557-9091}
\affiliation{%
  \institution{North Carolina State University}
  \city{Raleigh, North Carolina}
  \country{USA}}
\email{wguezas@ncsu.edu}

\author{Yongluan Zhou}
\orcid{0000-0002-7578-8117}
\affiliation{%
	\institution{University of Copenhagen}
	\city{Copenhagen}
	\country{Denmark}
}
\email{zhou@di.ku.dk}

\renewcommand{\shortauthors}{Laigner et al.}

\begin{abstract}

Microservices emerged as a popular architectural style over the last decade. Although microservices are designed to be self-contained, they must communicate to realize business capabilities, creating dependencies among their data and functionalities. Developers then resort to asynchronous, event-based communication to fulfill such dependencies while reducing coupling. 
However, developers are often oblivious to the inherent challenges of the asynchronous and event-based paradigm, leading to frustrations and ultimately making them reconsider the adoption of microservices. To make matters worse, there is a scarcity of literature on the practices and challenges of designing, implementing, testing, monitoring, and troubleshooting event-based microservices. 

To fill this gap, this paper provides the first comprehensive characterization of event management practices and challenges in microservices based on a repository mining study of over 8000 Stack Overflow questions. Moreover, 628 relevant questions were randomly sampled for an in-depth manual investigation of challenges.
We find that developers encounter many problems, including large event payloads, modeling event schemas, auditing event flows, and ordering constraints in processing events. This suggests that developers are not sufficiently served by state-of-the-practice technologies. We provide actionable implications to developers, technology providers, and researchers to advance event management in microservices.

\end{abstract}




\maketitle

\section{Introduction}
\label{sec:introduction}


The emergence of cloud computing as a paradigm for large-scale deployment of services has prompted industry practitioners to rethink how applications are architected and deployed~\cite{cloud}. In particular, we witness a growing adoption of \textit{microservice architectures}~\cite{microservices_tenets}. Microservice architectures promote designing components as independent building blocks that are deployed independently and interact with each other via network protocols~\cite{francesco:2017}. This architectural style brings more flexibility for the development and deployment of large systems, in contrast to conventional architectures, in which software components are strongly coupled with each other and deployed as a whole unit~\cite{newman:15}. 


The main benefit of microservices is that development teams can develop and operate their services independently~\cite{waseem2020systematic}. Microservices should be designed as self-contained software units that operate autonomously~\cite{newman:15}. To realize business capabilities, microservices need to communicate, which ends up creating dependencies among their data and functionalities~\cite{vldb}.
To manage such dependencies while maximizing decoupling, developers often rely on asynchronous event-based communication between microservices, instead of using traditional remote procedure calls~\cite{olep,LaignerKLSO20,OVEREEM2021110970}. The rationale is that, through events, producers and consumers are not directly coupled. For instance, producer microservices can trigger operations in other microservices or communicate their own state updates through events. 
Such an approach may positively influence the application's ability to evolve over time, limit the propagation of faults, and favor increased scalability of individual components~\cite{base,olep}.

Event-driven architecture has been rapidly gaining industry popularity~\cite{cabane2024impact}. However, despite the benefits of an event-driven design, developers frequently report challenges in managing events in industrial settings~\cite{uber_microservice,uber_proxy,uber_payment, uber_money,nubank,nubank-arc,airbnb_double,airbnb_integrity,wix_pitfalls,wix_kafka,OVEREEM2021110970,vldb,LaignerKLSO20}. 
Not surprisingly, it is common to encounter developers seeking support on how to implement event-driven microservices appropriately, such as this comment found on Stack Overflow (a popular Questions\&Answers forum)~\cite{quote_intro}: \textit{``Implementing an eventually consistent distributed architecture has turned out to be a pain. There are tons of blog posts telling stories about how to do it, but not showing (code) how to actually do it. One of the aspects I'm suffering is having to deal with manual retries of the messages when they haven't been acknowledged.''}
To make matters worse, in our study, we observed that there is a scarcity of literature on the practices and challenges that the adoption of an asynchronous paradigm brings to the designing, implementing, testing, monitoring, and troubleshooting of microservices. The challenges of managing communication in asynchronous and event-based systems have been explored in the fields of databases and distributed systems~\cite{flink,synapse}, but this knowledge is not widely disseminated among developers~\cite{tackling,vldb}, making them sometimes try to ``reinvent the wheel'' when addressing common challenges. This limited understanding of existing practices and how to address challenges in architectures based on asynchronous communication leads to frustrations and ultimately makes companies
reconsider the adoption of microservices.\footnote{Recent empirical studies have reported cases of companies moving from microservices back to monoliths~\cite{mendoncca2021monolith,su2024microservice}.}




There has been extensive work investigating the adoption and implications of microservice architectures \cite{vldb,zhou2021fault,ramirez2021empirical, waseem2023understanding, michael2023empirical, wang2021promises, vale2022designing, di2018migrating, zhang2019microservice, nasab2023empirical,bandeira2019we}. However, these pieces of work mostly focus on REST (representational state transfer) architectural style, describe the benefits of using microservices, or characterize data management and security practices without a holistic view of the practices and challenges. Furthermore, studies target specific companies or a limited number of practitioners, reducing the generalizability of their findings. 
To the best of our knowledge, no work has systematically investigated the practice and challenges brought about by an \emph{event-driven} design in microservice architectures. 
As a result, although \emph{event-driven} microservices form an important portion of microservice deployments today~\cite{vldb,waseem2023understanding}, it is unclear what particular challenges developers face when developing and operating them. The need to address these issues is even more relevant when we consider studies suggesting there is a tension between the use of asynchronous microservice designs and application safety~\cite{vldb}.


This work aims to characterize the practices and challenges faced by developers when adopting event-based microservice architectures.
More specifically, we first investigate the practices adopted by developers to manage events in microservice architectures. We aim to comprehend current technological trends and common implementation patterns related to the adoption of event management. We also seek to identify the functional and non-functional requirements that are met or supported by implementing event management in microservices. Secondly, we want to identify recurrent challenges faced by developers when managing events in microservice architectures. Focusing on these two perspectives, we can equip developers with knowledge of existing practices for event management in microservices and make them aware of the challenges they will face when developing asynchronous and event-based microservices.

To achieve our goal, we perform, to the best of our knowledge, the first empirical study to comprehend the practice and identify the challenges in managing events in microservice architectures from developers' perspective. 
To this end, our study is based on repository mining~\cite{gold2022ethics}. We mine and analyze relevant questions from Stack Overflow (SO),\footnote{\url{https://stackoverflow.com/}} one of the most popular Questions\&Answers forums for developers who seek technical advice or assistance~\cite{chen2016mining}. 
We collect more than 8,000 SO questions associated with managing events in microservice architectures, ensuring the diversity, representativeness, and quality of our dataset. Via a mix of keyword analysis and manual examination of tags, questions, and related responses (i.e., posts), we collect the most popular patterns and non-functional and functional requirements mentioned in the context of event management in microservices.
Then, we randomly sample 624 relevant SO questions for manual analysis to identify the challenges. We carefully extract the challenges behind each question through a peer-reviewed process. We classify the uncovered challenges into different properties of distributed systems and functional and non-functional requirements. These ultimately represent key areas that microservice developers struggle with. 

In the quantitative analysis, we observe that event management has been gaining increasing attention in practice, indicating the emergence of this architectural paradigm and the timeliness of this study.
Results show the popularity of patterns and requirements related to achieving data consistency and loose coupling. This suggests microservice developers seek to strike a better balance between consistency and decoupling in their microservice architectures by employing asynchronous events, especially in computations that span microservices. 
Furthermore, we observe that developers report different categories of patterns in their implementations. For instance, observability (e.g., \textit{distributed tracing}), performance (e.g., \textit{circuit breaker}), and security (e.g.,  \textit{access token}) patterns are reported being applied in questions involving event management. 

In the qualitative analysis, we find that microservice practitioners face a myriad of challenges in managing events. While events are supposed to maximize the decoupling of microservices~\cite{LaignerKLSO20,fowler:06}, developers often surprisingly necessitate synchronizing events for correct event processing and coordinating microservices to allow for software evolution, contradicting the alleged benefits of an event-based architecture~\cite{fowler:17,base}. Besides, although events can serve as natural progress markers of microservices~\cite{olep,fowler:17}, we observe that developers find little benefit in using events to monitor and troubleshoot microservices. In particular, developers have a hard time tracking down the result of their computations, which often span a network of dependent microservices and exhibit additional difficulties in replaying past events for debugging purposes.
Safeguarding security properties is also an emerging challenge in event-based microservice architectures. Developers struggle to synchronize event management and security technologies to ensure that only events generated by secured channels are processed and that only authenticated microservices can consume them.

This paper's contributions are manifold. We evidence that microservice developers are insufficiently served by state-of-the-practice technologies, including messaging systems, frameworks, databases, and cloud providers. They end up implementing their own ad hoc solutions to fulfill their requirements, which ultimately leads to errors and bugs, rendering adoption of a microservice architecture frustrating. 
Characterizing this emerging practice well informs general software developers, especially those unfamiliar with the challenges of designing distributed systems, about the hidden dangers of asynchronous, event-based microservice designs. We derive actionable implications for each class of event management challenges to researchers and message, database, cloud, and framework providers. These relate but are not limited to, devising novel microservice programming models, event management architectures, tools, and methods to advance the state-of-the-art practice in managing events in microservice architectures, making this a timely and valuable study.
Furthermore, the results of our work can benefit different stakeholders involved in the development and management of event-driven microservice architectures:
\begin{itemize}
    \item \textit{For practitioners:} the challenges reinforce the importance of reliable sources of information and the need for careful analysis of the documentation of the technologies they adopt. Practitioners should question gray literature with general and high-level instructions that often do not embrace the complex nature of problems faced in production settings. Practitioners should not overlook problem-prevention measures, which include running, testing, and deploying their applications under the expected workload prior to production. Besides, developers must be cautious about ``reinventing-the-wheel'' approaches, since for every possible design attempt, there are reports of previous attempts. 

    \item \textit{For researchers:} software engineering, system engineering, and database communities must work closely to each other. Leveraging the knowledge from other fields can help developers address their challenges. Software engineering researchers may find no appropriate solutions for their problems in tools from system engineering, requiring only appropriate interfaces while safeguarding the application. Also, researchers must embrace/meet practitioners' expectations that were analyzed and discussed in the paper, solving real-world problems without focusing on ``in vitro'' experiments. 

    \item \textit{For tool builders:} documentation must pair up with/highlight the problems faced in practice. They should also provide actions that can be taken to mitigate the shortcomings of the existing tools. Furthermore, vendors must be clear about what requirements and guarantees cannot be properly achieved.
        
    \item \textit{For educators:} we advise incorporating distributed systems concepts in software engineering courses, in particular algorithms and systems guarantees and their relationship with modern application development in the cloud. Yet, training students on how event management relates and makes its presence in modern application architectures, such as serverless, microservices, SaaS, to name some. 
\end{itemize}

In addition, the dataset used in this
study is made available~\footnote{\url{https://zenodo.org/records/13149520}} as an additional contribution to the research community, allowing other researchers to further investigate the theme.


The remainder of this work is organized as follows: Section 2 describes the role of events in modern microservice architectures and typical issues developers encounter. In Section 3, we present the methodology employed in this work. Section 4 presents the state of practice. Section 5 presents the specific challenges microservice developers encounter in managing events and their implications for the research community and industry providers. Section 6 discusses the threats to validity. Section 7 presents the related work. Lastly, Section 8 concludes this work.


\section{Background}
\label{sec:background}



\begin{figure}
  \includegraphics[width=1\textwidth]{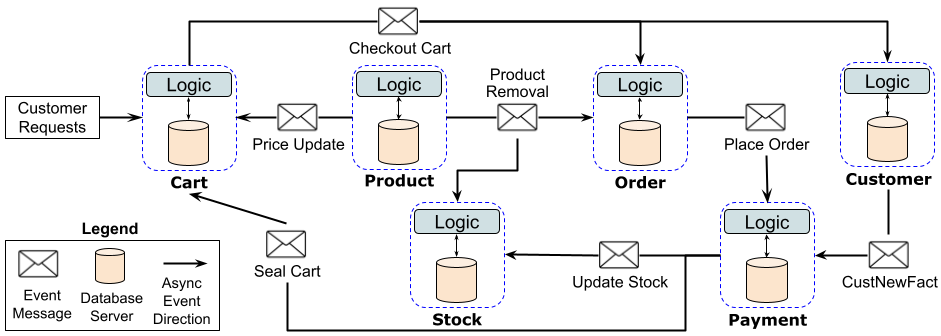}
  \caption{Events in a microservice architecture}
  \Description{Events in a microservice architecture}
  \label{fig:microservice}
  \vspace{-3ex}
\end{figure}

To discuss the role events play in microservice architectures, we use Figure~\ref{fig:microservice} to illustrate an ecommerce platform. We base on a popular microservice open-source project~\cite{eShopOnContainers} that incorporates real-world event processing patterns.

\subsection{Context}

Upon a customer's checkout request, the \textit{Cart} microservice spawns the execution of a checkout through generating a \texttt{checkout\_cart} event. The event is then consumed by the \textit{Order} microservice, which processes the cart items and generates the \texttt{place\_order} event. Next, the \textit{Payment} microservice is responsible for processing the payment and generating the \texttt{update\_stock} and \texttt{seal\_cart} events for downstream consumption by the \textit{Stock} and \textit{Cart} microservices, respectively.

It is noteworthy that, in synchronous communication paradigms, such as through REST APIs and RPCs, the requester microservice gets blocked until the request terminates in the remote microservice and the response is received. It is natural to deduce that such blocking times can grow arbitrarily when business transactions traverse several microservices. In the example above, \textit{Cart} would necessarily wait until \textit{Payment} terminates the payment processing. Through the asynchronous paradigm, though, once the \texttt{checkout\_cart} event is generated, \textit{Cart} microservice is free to allocate computational resources (e.g., threads, CPUs, and memory) to serve other requests. This usually favors higher efficiency in event processing 
~\cite{hal-04112339}.

As exhibited in Listing 1, the communication through events is only possible because events carry semantic data that is used by microservices in distinct ways~\cite{fowler:17}. For instance, although the payment data is sent by the customer in a checkout request, such data is only used by the \textit{Payment} microservice. That requires both \textit{Cart} and \textit{Order} microservices to pass along payment data in the \texttt{checkout\_cart} and \texttt{place\_order} events, respectively.


\begin{lstlisting}[caption={Example of application-generated event types},xleftmargin=.25\textwidth,label={lst:event_types},language=rock,captionpos=b,basicstyle={\footnotesize\ttfamily\linespread{0.1}}]]
product_update:{ product_id, type, description, ... }
price_update:{ product_id, old_price, new_price, ... }
checkout_cart:{ customer_id, items:[{ product_id, qty,  .. }], 
    card_num, exp_date, address, discount }
update_stock:{ product_id, qty, timestamp, ... }
\end{lstlisting}

\subsection{Problem Statement}

Besides the apparent attractiveness of a decoupled design, managing events appropriately renders challenges. As events are generated asynchronously, that is, producers do not wait for consumers' reception, developers must account for the anomalies that possibly arise in event delivery, as described by the microservice adopter Nubank~\cite{nubank-arc}:


\begin{displayquote}
``you have late-arriving events, you can make a payment that we receive that should be credited on Friday, we need to time travel back, re-play the events and figuring out the balance that does not invalidate the invariants we have''
\end{displayquote}

Besides, it is often reported that developers necessitate making sure anomalies in the event order does not lead to problems in their microservices, as explained by an Uber team~\cite{uber_money}:

\begin{displayquote}
``If it consumes an event that is not in sequence, our processing logic identifies the version mismatch and we retry the event a number of times.''
\end{displayquote}


Turning to our example scenario, if \texttt{price\_update} events are processed in different orders, that can lead to incorrect product prices. Besides, supposing that the \texttt{product\_removal} is not delivered to and processed by both \textit{Order} and \textit{Stock}, that leads the system to an incorrect state. In addition, if \texttt{place\_order} is delivered more than once, that can lead to overcharging customers. An abnormal burst of events can also pose challenges to the system if consumers cannot cope with the arrival rate of events. In this work, we delve into these and several other types of challenges that affect event management in microservice architectures.


\section{Study Design}

Although event-driven microservices are a popular architectural style in industry settings~\cite{vldb,base,olep}, existing studies mostly outlook the practical implications of adoption and the challenges brought about by an asynchronous and event-based design to microservices. Thus, the goal of this work is to \textit{identify the practices adopted and challenges faced by practitioners developing software systems using event-driven microservice architectures.}

To achieve this goal, our study focus on answering two research question, as follows:

\begin{itemize}
    \item \textbf{RQ1. What is the state of the practice on managing events in microservice architectures?}
    \textit{Rationale}: Despite the popularity of event-driven microservice architectures, in the literature it is little known on why practitioners adopt this architectural style. Also, the literature is scarce on describing what are the practices adopted by them to manage events in microservice architectures, allowing practitioners to understand technological trends and popular implementation patterns associated with event management adoption. Thus, with this RQ, the insights we want to gain is threefold: (i) Understand the popularity trend of event management in microservice architectures; (ii) Collect the most popular patterns associated with event management adoption; (iii) Identify what functional and non-functional requirements are either enabled or facilitated by managing events in microservices.
    
    \item \textbf{RQ2. What challenges do practitioners encounter when managing events in microservice architectures?} \textit{Rationale}: Once we understand the current practices on developing software systems using event-driven microservice architectures, we focus on identifying what particular challenges developers face when developing and operating them. In this question, we delve into the specific issues and impediments microservice developers face in practice while managing events.
\end{itemize}

To answer these two research questions, we use Stack Overflow (SO)~\cite{stack}, a popular question-and-answer online platform where practitioners seek technical assistance on issues they face in their day-to-day activities to understand the challenges of managing events in microservice architectures.
Our methodology is similar to the ones adopted in previous studies~\cite{bagherzadeh2019going,yang2016security,wen2021empirical} to collect, filter, and analyze questions. However, our focus is on event management in microservice architectures.
We detail the procedure of our study below.


\subsection{Data Collection}



\subsubsection{Downloading Stack Overflow dataset}

We started by downloading the entire Stack Overflow dataset $S_{all}$ from the official Stack Exchange Data Dump~\cite{stackexchange2023} available when we started this study (September 15, 2023). The dataset includes 23.199.461 questions dated from July 31, 2008 to December 31, 2022.\footnote{We excluded the year 2023 from our analysis as it was still a year in progress when this work was carried out} Every question in the dataset has title, body and a ``tag'' metadata, which denotes the topics on which the question lies.



\subsubsection{Exploring initial tag set}

We start with a general tag set to include as many relevant questions as possible. Thus, we use $T_{ini}$ = \{``microservice''\}, resulting in 8369 questions, from which we performed an exploratory search. We observed that many questions were unrelated to event management, requiring significant effort to filter out. Also, we observed that relevant questions would mention the keywords ``microservice'' and ``event'' in either the title or body.

\subsubsection{Filtering by relevant keywords}

With the above insight in mind, we filtered the questions in $T_{ini}$ that contain the keywords ``microservice'' and ``event'' in their title or body, leading to a total of 1407 questions (denoted as $S_{rel}$). 
We extracted 566 tags from $S_{rel}$, which we denote as $T_{key}$. Table~\ref{tab:tags} shows the distribution of questions per tag in $S_{rel}$ (only the tags with five questions or more). 

The first two authors jointly examined the candidate tags and observed that many of them relates to techniques, patterns, and technologies applied to managing events in microservice architectures, such as \textit{event-driven architecture}, \textit{pubsub}, \textit{message queue}, and \textit{event sourcing}. Even tags unrelated to event management (e.g., \textit{authentication}, \textit{logging}, and \textit{database}) were often accompanied by event management-related tags, indicating possible correlated issues. Both observations gave us the confidence to proceed with further analysis.



\begin{table}
 \caption{Relevant tags extracted from Stack Overflow}
 \centering
    \begin{tabular}{p{50mm}c|p{50mm}c}
     \hline
        \textbf{Tag} & \textbf{$\#$Questions} & \textbf{Tag} & \textbf{$\#$Questions} \\
        \hline
        \hline
        microservice & 895 & messaging, saga & 22 \\
        \hline
        architecture & 124 & database & 20 \\
        \hline
        event-sourcing & 90 & masstransit, publish-subscribe & 18 \\
        \hline
        apache-kafka & 78 & event-handling, asynchronous & 17 \\
        \hline
        domain-driven-design & 76 & eventual-consistency, soa, distributed-transactions & 15 \\
        \hline
        cqrs & 71 & vert.x, distributed-computing & 11 \\
        \hline
        rabbitmq & 54 & distributed-system, authentication & 10\\
        \hline
        event-driven & 43 & redis, google-cloud-pubsub & 9 \\
        \hline
        events & 38 & azure-service-fabric, event-bus, messagebroker, authorization & 8 \\
        \hline
        event-driven-design & 36 & spring-cloud-stream, system-design, websocket, azureservicebus, socket.io, web-services & 7 \\
        \hline
        design-patterns & 31 & spring-kafka, bounded-contexts, database-design, event-based-programming, aws-lambda, google-cloud-platform, transactions, logging, amqp & 6 \\
        \hline
        message-queue & 23 & apache-kafka-streams, software-design, security, identityserver4, integration, <service> & 5 \\
        \hline
      \end{tabular}
    \label{tab:tags}
\end{table}

\subsection{Analyzing the State of the Practice}



\subsubsection{Patterns Trend}

Based on a popular collection of patterns for microservice architectures~\cite{patterns}, we extracted from the questions in $S_{rel}$ the patterns for microservices mentioned by developers. It noteworthy we also included in the analysis the different users' posts (i.e., responses) for each question, totaling 3142 entries.
We took extra care analyzing the dataset to embrace all possible forms of writing the same pattern (including typos), synonyms, and acronyms. For instance, we also identified the pattern "Command Query Responsibility Segregation" (CQRS) via "cqrs" or "CQRS." Section~\ref{subsec:patterns} answers RQ1(b).


\subsubsection{Non-functional Requirements Trend}

We also analyzed the most common non-functional requirements (NFRs) mentioned by practitioners.
We use as reference the NFRs listed in Chung et al.~\cite{chung2012non}. We started searching keywords related to NFRs terms in $S_{rel}$ (including posts) and incrementally introduced acronyms and synonyms.
After introducing each keyword, we evaluated randomly selected related questions to validate whether the classification of the NFR fit the questions' context. We continued this process until exhausting the set of possible NFRs.

We understand that an NFR keyword search does not necessarily mean that such a requirement is part of the practitioners' problem. However, our analysis observed that such cases are often exceptions rather than rules, thus not undermining the analysis of the most pursued NFRs regarding event management in microservices. We respond to RQ1(c) in Section~\ref{subsec:nfr}.

\subsubsection{Functional Requirements Trend}

We were also interested in understanding the specific functional requirements pursued by practitioners that are either enabled by or facilitated by managing events. Unlike NFRs, though, we noticed that uncovering functional requirements often necessitated a thorough, detailed analysis to uncover the application scenario and, consequently, the functional requirement related to event management.

\noindent\textbf{\textit{Relevance Filter}.} Given the high number of questions, manually examining all of them would demand a substantial amount of time, which could make reporting this study's evidence in a timely fashion impracticable. Therefore, at this point, we decided to apply a relevance filter.
First, we use a relevance heuristic $\nu$ to filter out questions with tags that are employed to a much lesser extent compared to others. As previous studies~\cite{bagherzadeh2019going,wen2021empirical}, we considered only the tags whose $\nu$ is higher than 0.005, leading to a set of 1298 questions and 93 associated tags.

\begin{equation*}
\nu = \frac{\# \text{ of questions with tag t in }S_{rel}}{\# \text{ of questions in }S_{rel}}
\end{equation*}

Next, we focused on prioritizing the analysis of questions tagged with event management technologies (e.g., rabbitmq, kafka, masstransit) and techniques (e.g., cqrs, event-sourcing, event-handling) rather than questions tagged with technologies not directly related to event management (e.g., python, java, php, elasticsearch). Although the former also presents problems in the event management domain, thus part of our problem scope, we observe filtering by related event management tags led more often to questions revealing challenges within our study scope. 

These two filters combined led to 53 tags (shown in Table~\ref{tab:tags}) with a total of 925 questions. To proceed with the manual examination, we randomly sample 628 questions, which represents more than 2/3 of the relevant questions filtered.

\noindent\textbf{\textit{Procedure}.} We adopt an open coding procedure~\cite{open_coding} to analyze the sampled questions to pinpoint the specific functional requirements sought by practitioners inductively.

The first two authors, both of whom have years of software and data engineering experience, jointly participate in the manual examination of the questions. They analyze the sampled questions multiple times to familiarize themselves with them. In this process, many elements of the question were taken into consideration for inspection, including the title, body, code snippets, URLs, and the author's responses to other users' inquiries, which can ultimately clarify unclear points of the original problem statement (i.e., content in the body). 

We observed that the user explicitly expressed the functional requirements being sought in most of the questions. For instance, in data replication cases, users would often mention that an event is generated based on some update (e.g., a user credit card score), so other microservices can also apply this change to their databases (see \textbf{S3} in Section~\ref{subsec:scenarios}).

However, in other cases, the specific functional requirement is not made explicit. For instance, different terms, like "consistency," are used to denote replicating data as a functional requirement, like the example as follows~\cite{quote_53100208}:

\begin{displayquote}
\textit{Customer} and \textit{Order} microservice both have customer details, though in \textit{Order} microservice customer infos are striped to only required fields. I understand there is a way to maintain \textbf{consistency of data across microservices using events}.
\end{displayquote}

Therefore, upon clarifying the functional requirement, the authors give terms to represent the requirements of the questions, using the terms used by developers whenever possible. We also found cases where a question contains more than a functional requirement. This way, we count the number of functional requirements independently of the number of questions. Section \ref{subsec:scenarios} answers RQ1(d).

\subsection{Characterizing the Challenges}


We use the same randomly sampled question set used above to analyze specific issues that developers report and characterize the challenges of managing events in microservice architectures.

Again, the first two authors adopt an open coding procedure and analyze all the elements of each question, including comments from users other than the author's question, to carefully extract insights about practitioners' challenges. The detailed procedure is as follows.

First, we ensured the practitioner's requirements were properly clarified. The reasoning is that we noticed the user often reports a problem faced subsequent to or in the context of expressing the non-functional and functional requirements sought. Since we already had the questions labeled with requirements through the previous methodology steps, our primary task remained to understand the challenges from the problem description and other question elements.

In many cases, the challenges could be identified by questions raised in the form of "how" or "what," as in other studies~\cite{wen2021empirical}. For example, one user asked~\cite{quote_29148547}: "How can I process the event with the user credentials?" 
; and another user inquired~\cite{quote_49986434}: "What's maybe not a good idea in trying to recover the current state of your domain model by replaying an arbitrary set of your events 
."

We also observed, though, in other cases, the users would not raise a question, but rather: \newline
\textit{(a)} express the willingness to achieve certain functionality~\cite{quote_57328269}: "I want to create a third microservice that is responsible to join the data of \textit{ProductService} and \textit{StoreService}." 
\newline
\textit{(b)} describe the possible cause of a problem~\cite{quote_71971389}: "In my system two different sources can cause creating specific type of event. [...] due to replication lag" 
\newline
\textit{(c)} inform a desired correctness criteria not being met~\cite{quote_67612615}: "The end price should be 100 for that product, but sometimes these events are processed in random order." 

In a few cases, the question's body lacks detailed information about the user's application scenario. In those cases, we searched for additional comments from the author's post made as a response to questions raised by other SO users. In the absence of further details about the user's problem scenario, the authors made the best effort to characterize the reported issue, jointly discussing whenever a disagreement was in place.

Second, once the problem statement was clarified, the authors started grouping similar problems into categories. The authors jointly iterated multiple times over the questions and categories. Whenever conflicts were observed in grouping the questions, a third arbitrator was introduced to discuss and reach a consensus. The third arbitrator has more than fifteen years of experience in cloud computing and data engineering. Lately, all questions come to an agreement and the final categories (i.e., the challenges) are confirmed by all the participants. Section \ref{sec:challenges} answers RQ2(a).

\section{State of the Practice (RQ1)}
\label{sec:state_of_practice}

In this section, we describe the state of practice based on the discussions found in SO. 

\subsection{Patterns Trend}
\label{subsec:patterns}

Table~\ref{tab:patterns} exhibits the most recurrent patterns that appear in questions related to event management in microservices architectures~\footnote{For improved presentation, we show only those patterns appearing in 20 or more questions.}. In the top ten most mentioned patterns, four are directly related to event management (\textit{messaging}, \textit{event sourcing}, \textit{domain event}, \textit{cqrs}), one related to synchronous communication (RPC), two related to deployment patterns (\textit{database per service} and \textit{service-per-container}), and three related to typical microservice architectural patterns (\textit{API gateway}, \textit{service registry}, and \textit{aggregate}).

We can observe that although event management patterns dominate the list, patterns unrelated to event management also appear substantially. For instance, remote procedure invocation, a synchronous communication technology, appears as the second most cited pattern. As events are processed asynchronously, the employment of synchronous communication mechanisms contrasts with the pursued benefits of events, which are often related to more efficient usage of computational resources and decoupling~\cite{LaignerKLSO20,olep,fowler:17}. This trend is mainly due to the need for developers to reply to users the result of asynchronous computations, as exemplified in a developer's quote below:

\begin{displayquote}
``How is it possible to send the results stored in the kafka topic back to the requesting WebClient\_X?''
\end{displayquote}

These challenges are further discussed in Section~\ref{subsec:safety}. Popular microservice patterns appearance in the top 10, namely, \textit{API gateway} and \textit{service registry}, can be justified by the need to allow web clients to initiate asynchronous computations and expose certain APIs to the external world, and to allow for service discovery
, respectively. It is worth noting these are not concerns event management tackles in an event-driven architecture.

We also observed that the patterns \textit{Database per Service} and \textit{Service-per-Container} are often used synonymously by developers. The popularity of these two database deployment patterns suggests developers tend to avoid deployments that prescribe two or more microservices sharing the same database. For instance, the \textit{messaging} pattern is only mentioned together with \textit{Shared Database} is 12 occasions. We further discuss this trend in Section~\ref{subsec:nfr}.

\vspace{1ex}
\begin{mdframed}
\textbf{Highlight:} Adopting a shared database pattern would jeopardize the benefits of decoupling through events. The microservices would compete for computational resources of the machine (or container) hosting the database server, leading to decreased performance effects.
\end{mdframed}
\vspace{1ex}

To aid our analysis, we grouped the patterns into pairs and count the number of times they are cited together in questions, as exhibited in Table~\ref{tab:pattern_pair}. In general, the pairs' appearance is aligned with the individual numbers of Table~\ref{tab:patterns}. Event management patterns, such as \textit{messaging}, \textit{event sourcing}, \textit{CQRS}, and \textit{domain event}, appear substantially is pairs followed by \textit{Remote Procedure Invocation} and \textit{Database Per Service}. Other interesting insights are also confirmed, like the popularity of an authentication mechanism together with service discovery.
Furthermore, the combination of pairs sheds light on the most common set of patterns considered in questions involving event management in microservices. \textit{Messaging}, \textit{event sourcing}, \textit{Database per Service}, \textit{CQRS}, \textit{Service Registry}, and \textit{Third Party Registration} are substantially mentioned in conjunction in questions, suggesting a pattern adoption trend.

\begin{figure}
  \centering
  \includegraphics[width=0.8\textwidth]{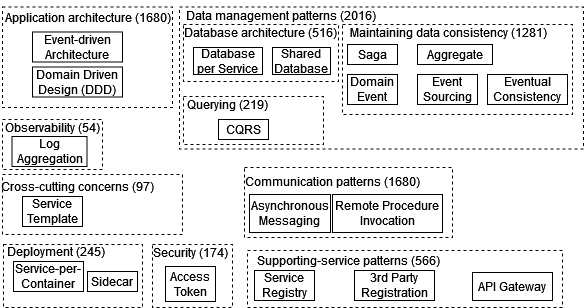}
  \Description{Most popular patterns in event management in microservices}
  \caption{Most popular patterns in event management in microservices}
  \label{fig:pic_patterns}
\end{figure}

To further aid our analysis, we grouped correlated patterns into categories shown in Figure~\ref{fig:pic_patterns}. Patterns associated with data management form the most prominent category, followed by communication patterns, application architecture, support-service, deployment, security, and observability. We observe three subgroups within data management, including patterns associated with data consistency enforcement, database deployment, and data querying.
Interestingly, in data consistency, we observe not only event-related patterns, such as \textit{domain event} and \textit{event sourcing}, but also \textit{SAGA} and \textit{eventual consistency}. That suggests developers also look for event management as an alternative to traditional, synchronous-based mechanisms for implementing consistency patterns, confirming the preliminary findings of~\cite{vldb}.

\vspace{1ex}
\begin{mdframed}
\textbf{Highlight:} Asynchronous events can enhance the performance of data consistency patterns by allowing for non-blocking interactions across microservices. That can lead to higher performance.
\end{mdframed}
\vspace{1ex}

In the application architecture realm, event-driven architecture (EDA) is substantially mentioned, an expected trend given the scope of the study. However, Domain-Driven Design (DDD) also appears to a lesser extent. We found that only twelve questions mention EDA and DDD in conjunction, suggesting these two patterns are not popularly adopted together when it comes to event management in microservices.

\vspace{1ex}
\begin{mdframed}
\textbf{Highlight:} As EDA often prescribes events trigger specific functions~\cite{fowler:17}, our analysis suggests that this direct mapping between an event type and an application function (aka business logic) demotivates the modeling of extra modularity layers as prescribed by DDD.
\end{mdframed}
\vspace{1ex}

We also observe the representativeness of patterns associated with deployment, such as \textit{sidecar} and \textit{service-per-container}, security with \textit{access token}, and observability with \textit{log aggregation}. The categories and related patterns exhibited in Figure~\ref{fig:pic_patterns} form the bulk of patterns cited by microservice developers. 
We explore further these phenomena in Section~\ref{sec:challenges}.

\begin{table}
 \caption{Relevant patterns extracted from Stack Overflow}
\centering
      \begin{tabular}{c|c|c|c}
        \hline
        \textbf{Pattern} & \textbf{$\#$ Posts} & \textbf{Pattern} & \textbf{$\#$ Posts} \\
        \hline
        \hline
        Messaging & 1056 & Shared Database & 50 \\
        \hline
        Remote Procedure Invocation & 624 & Circuit Breaker & 48 \\
        \hline
        Event Sourcing  & 537 & Transactional Outbox & 43 \\
        \hline
        Database per Service & 424 & Domain-specific & 42 \\
        \hline
        API Gateway & 249 & Single Service per Host & 39 \\
        \hline
        Domain Event & 225 & Service-per-VM & 38 \\
        \hline
        Service Registry & 221 & CDC & 38 \\
        \hline
        Command Query Responsibility Segregation (CQRS) & 219 & Backend for frontend & 32 \\
        \hline
        Aggregate & 207 & Service per Team & 32 \\
        \hline
        Service-per-Container & 192 & Application Metrics & 25 \\
        \hline
        Access Token & 174 & Health Check API & 25 \\
        \hline
        Eventual Consistency & 171 & Externalized Configuration & 24 \\        
        \hline
        Event-driven Architecture & 151 & Polling Publisher & 23 \\
        \hline
        Saga & 141 & Distributed Tracing & 23 \\
        \hline
        Domain Driven Design (DDD) & 138 & Self-contained Service & 23 \\  
        \hline
        Service Template & 97 & Audit Logging & 20 \\
        \hline
        3rd Party Registration & 96 & Transaction Log Tailing & 20 \\
        \hline
        Log Aggregation & 54 & Multiple Service per Host & 20 \\
        \hline
        Sidecar & 53 \\
        \hline
      \end{tabular}
  \label{tab:patterns}
\end{table}

\vspace{1ex}
\begin{mdframed}
\textbf{Finding 1:} There is a specific subset of patterns that appear frequently in event management questions. These mainly relate to database resource isolation, data consistency, and asynchronous messaging.
Questions involving event management are also accompanied by mentions of patterns not directly related to event management, suggesting users often attempt to blend heterogeneous patterns in their deployments to solve cross-cutting concerns such as observability and security.
\end{mdframed}
\vspace{1ex}

\begin{table}
\caption{Relevant pairs of patterns mentioned in event management extracted from Stack Overflow}
  \begin{tabular}{c|c|c|c}
    \hline
    \textbf{Pair of Patterns} & \textbf{$\#$ of Posts} & \textbf{Pair of Patterns} & \textbf{$\#$ of Posts} \\
    \hline
    \hline
    Remote Procedure Invocation \& Messaging & 278 & Service Registry \& Remote Procedure Invocation & 69 \\
    \hline
    Messaging \& Event Sourcing & 202 & Messaging \& CQRS & 68 \\
    \hline
    Remote Procedure Invocation \& Event Sourcing &	145 & Service-per-Container \& Remote Procedure Invocation & 67 \\
    \hline
    Messaging \& Database per Service &	132 & Messaging \& Aggregate & 67 \\
    \hline
    Remote Procedure Invocation \& Database per Service	& 125 & Remote Procedure Invocation \& Domain Event & 67 \\
    \hline
    Event Sourcing \& Database per Service & 124 & Messaging \& Eventual Consistency	& 64 \\
    \hline
    Event Sourcing \& CQRS	& 122 & SAGA \& Messaging & 60 \\
    \hline
    Service Registry \& Messaging & 114 & Domain Event \& CQRS & 58 \\
    \hline
    Service Registry \& 3rd Party Registration & 90 & Messaging \& Access Token & 57 \\
    \hline
    Event Sourcing \& Domain Event	& 85 & Domain Event \& DDD	& 56 \\
    \hline
    Messaging \& API Gateway & 84 & Remote Procedure 
    Invocation \& Event-driven Architecture & 53 \\
    \hline
    Remote Procedure Invocation \& API Gateway	& 81 & CQRS \& Aggregate &	51 \\
    \hline
    Event Sourcing \& Aggregate	& 81 & Event Sourcing \& API Gateway	& 51 \\
    \hline
    Messaging \& Domain Event & 81 & Eventual Consistency \& Event Sourcing & 50 \\
    \hline
    Database per Service \& CQRS & 72 & Service Template \& Remote Procedure Invocation & 50 \\
    \hline
    Service-per-Container \& Messaging & 72 & Remote Procedure Invocation \& CQRS	& 50 \\
    \hline
    Messaging \& Event-driven Architecture & 71 \\
  \hline
  \end{tabular}
\label{tab:pattern_pair}
\end{table}

\subsection{Non-Functional Requirements}
\label{subsec:nfr}

Table~\ref{tab:nfrq} exhibits the extracted non-functional requirements from SO. We discuss the results along with key correlations found among NFRs as follows.


\begin{table}
    \caption{Relevant Non-Functional requirements extracted from Stack Overflow}
      \begin{tabular}{c|c}
        \hline
        \textbf{Non-Functional Requirement} & \textbf{$\#$ of Posts} \\
        \hline
        \hline
        Consistency & 312 \\
        \hline
        Decoupling & 303 \\
        \hline
        Scalability  & 249  \\
        \hline
        Performance & 194 \\
        \hline
        Modularity & 170 \\
        \hline
        Traceability & 164 \\
        \hline
        Security & 154 \\
        \hline
        Fault Tolerance & 138 \\
        \hline
        Load Balancing & 109 \\
        \hline
      \end{tabular}
    \label{tab:nfrq}
\end{table}

\begin{table}
    \caption{Relevant Non-Functional Requirement pairs extracted from Stack Overflow}
      \begin{tabular}{c|c}
        \toprule
        \textbf{Pair of NFR} & \textbf{$\#$ of Questions} \\
        \hline
        Performance (24.23\%) \& Scalability (18.87\%) & 47 \\
        \hline
        Decoupling (15.51\%) \& Scalability (18.87\%) & 47 \\
        \hline
        Consistency (14.42\%) \& Decoupling (14.85\%)  & 45 \\
        \hline
        Consistency (12.5\%) \& Scalability (15.66\%)  & 39 \\
        \bottomrule
      \end{tabular}
    \label{tab:features_pair}
\end{table}

\noindent\textbf{Consistency and Decoupling.} Consistency appears to be the most cited NFR in event management questions in sync with the microservice patterns collected in the last section. A popular concern we noticed in this regard is the unmet expectation of certain events to arrive. For instance, one developer mentions a consistency problem observed whenever the delivery of events lags behind~\cite{quote_30583466}:


\begin{displayquote}
"When an \textbf{event isn't received} in the worker, the counters start to \textbf{"drift away" from the true} MySQL count.[...]"
\end{displayquote}

We observe that most developers express awareness that there is a natural delay to be expected. In a few cases, though, concerns over "how up to date" a given microservice is  are implicitly raised, as exemplified by the same developer, as follows: 

\begin{displayquote}[]
It is \textbf{expected} that there is a \textbf{consistency delay} for this type of data. How up-to-date the data is can even be figured out and included in the responses for stats data.
\end{displayquote}


\noindent We further discuss the tension between the eventual delivery of events and application correctness in Section~\ref{sec:challenges}. Furthermore, we found that decoupling is the NFR most correlated with consistency. Analyzing the posts in which consistency and decoupling are mentioned together (45 posts), we observed that developers often identify trade-offs in pursuing the two NFRs in conjunction. The following extracted SO thread exemplifies this trend. A developer presents its application requirements starting with decoupling concerns~\cite{quote_43950808}:

\begin{displayquote}
While each microservice generally will have its own data - certain \textbf{entities} are required to be \textbf{consistent across multiple services}. \newline
\end{displayquote}

\begin{displayquote}
\textbf{I do not want shared database} architecture, where a single DB manages the state across all the services. That \textbf{violates isolation and shared-nothing} principles.
\end{displayquote}

And then highlights a possible mechanism to ensure cross-microservice consistency~\cite{quote_43950808}:

\begin{displayquote}
"I do understand that, a microservice can publish an event when an entity is created, updated or deleted. All other microservices which are interested in this event can accordingly update the linked entities in their respective databases." 
\end{displayquote}

However, the developer realizes some inner drawbacks~\cite{quote_43950808}:

\begin{displayquote}
"however it leads to a lot of careful and \textbf{coordinated programming effort across the services}. Can Akka or any other framework solve this use case? How?"
\end{displayquote}

Another developer responds the thread acknowledging the tension between decoupling and consistency and the lack of principled solutions to the problem~\cite{quote_44748028}:

\begin{displayquote}
I think there are 2 main forces at play here:\newline
\textbf{decoupling} - that's why you have microservices in the first place and want a shared-nothing approach to data persistence\newline
\textbf{consistency} - if I understood correctly you're already fine with eventual consistency\newline
\textbf{I don't know of any framework to do it out of the box}, probably due to the many use-case specific trade-offs involved.
\end{displayquote}

\noindent\textbf{Performance and Scalability.} Consistency and decoupling NFRs lead the number of post appearances in SO. However, taken together, they are outperformed by the pair performance and scalability. We observe developers express concerns on varied performance topics. For instance, both developers below face a problem when the event processing rate exacerbates the processing capacity of consumer microservices. We delve into performance issues on managing events in Section~\ref{subsec:performance}.

\begin{displayquote}
(DEV\#1) Service A listens to a rabbit queue and sends http request to service B (which takes a couple of seconds). Both services scale based on the number of message in the queue.
The problem is that the \textbf{requests from A to B are not balanced}. [...]
That obviously causes \textbf{low performance and timeouts}.~\cite{quote_73349618}
\end{displayquote}

\begin{displayquote}
(DEV\#2) My problem is the queue. \textbf{I can't get} an easily\textbf{ scalable queue that guarantees ordering} of the messages. It actually guarantees "slightly out of order" with at-least once delivery [...]~\cite{quote_53270770}
\end{displayquote}

The same developer continues then:

\begin{displayquote}
"But it turns out that using this solution, it will destroy \textbf{performance} when events are produced at high rates (I can use a visibility timeout or other stuff, the result should be the same)."
\end{displayquote}


\noindent\textbf{Fault tolerance.} We also observe developers consider the adoption of event management technologies as a mechanism to increase fault tolerance in their microservice architectures. In the example below~\cite{quote_48912603}, a developer inquiries other users about whether Kafka can provide appropriate fault-tolerance support. We discuss about failures and their possible impact on microservices in Section~\ref{subsec:safety}.

\begin{displayquote}
"I am working in a project that starts creating \textbf{independent deployable services}. The service we are creating should be \textbf{resilient} with an \textbf{24/7 uptime}." 
\end{displayquote}

\begin{displayquote}
"In this case Kafka will be used as an \textbf{event system}. What do you think about the requirements and the usage of \textbf{Kafka to get a highly available and resilient} application?"
\end{displayquote}

\vspace{1ex}
\begin{mdframed}
\textbf{Finding 2:} Consistency, decoupling, and performance appear as the most mentioned non-functional requirements. However, developers suggest there are trade-offs on properly meeting them in event-driven microservice architectures.
\end{mdframed}
\vspace{1ex}

\subsection{Functional Requirements}
\label{subsec:scenarios}

In this section, we discuss the most recurrent functional requirements sought by microservice developers on which event management is applied (denoted by \textbf{FR}[\textbf{0-N}]). Table~\ref{tab:scenario_questions} summarizes the number of functional requirements per question.

\begin{wraptable}{r}{8cm}
    \caption{Relevant Functional Requirements extracted from Stack Overflow}
      \begin{tabular}{c|c}
        \hline
        \textbf{Functional Requirement} & \textbf{$\#$ of Questions} \\
        \hline
        \hline
        FR1. Propagation of state updates & 78 \\
        \hline
        FR2. Multi-microservice workflows & 66 \\
        \hline
        FR3. Data Integrity Maintenance & 33 \\
        \hline
        FR4. Replaying of events & 22 \\
        \hline
        FR5. Query Processing & 20 \\
        \hline
        FR6. Data replication & 17  \\
        \hline
        FR7. Cache management & 12 \\
        \hline
        FR8. Task Scheduling & 5 \\
        \hline
      \end{tabular}
  \label{tab:scenario_questions}
\end{wraptable}

\noindent\textbf{FR1. Propagation of state updates.} We observe that the majority of questions contains a requirement related to propagating state updates in form of events. A developer summarizes this practice as follows~\cite{quote_43950808}:

\begin{displayquote}
"A microservice can publish an event when an entity is \textbf{created, updated or deleted}. All other microservices which are interested in this event can accordingly \textbf{update the linked entities} in their respective databases."
\end{displayquote}

Here we only take into account the questions where the developers' intention of propagating the events are either not expressed or not clear from the discussions. However, it is natural to deduce these state updates in form of events are used by other microservices to achieve other requirements, as we discuss next.



\noindent\textbf{FR2. Multi-microservice workflows.} We also observe a substantial number of developers reporting use cases where a business transaction require a composition of microservices~\cite{saga_io} via events. 

For instance, question 42140285 exhibits that an event generated by a microservice triggers an operation in another microservice~\cite{quote_42140285}:
\begin{displayquote}
"\textbf{\textit{Order} [microservice]} receives an order request. It has to \textbf{store} the new Order ([record]) \textbf{in its database} and \textbf{publish a message} so that \textbf{\textit{Payment} service} realizes it has to charge for the item"
\end{displayquote}

Differently from the previous scenario though,
we observe developers in this case have the expectation about the completion of the triggered computations.

In sum, in this functional requirement, the generation of an event act as a command, a call for action, for downstream microservices, manifesting the need to perform operations, often involving the operations in their private states.

\noindent\textbf{FR3. Data Integrity Maintenance.}

We also found questions that developers highlight how an event is used to maintain the integrity of a microservice's state. An example is provided as follows~\cite{quote_43378165}.

\begin{displayquote}
"I have a web-api-endpoint that receives orders that an \textit{OrderMS} is responsible to handle. When order is put[,] Inventory must be updated so \textbf{\textit{OrderMS}} publish an event to subscribers [...] and \textbf{\textit{InventoryMS} will update the inventory} due to it is subscribing to current event/message"
\end{displayquote}

Another developer explains the intention to achieve functional dependency across microservices via events propagated~\cite{quote_53100208}:
\begin{displayquote}
"Customer and Order microservice both have customer details, though in Order microservice customer infos are striped to only required fields. I understand there is a way to maintain consistency of data across microservices using events."
\end{displayquote}

\noindent\textbf{FR4. Replaying of events.} We observe that developers seek replaying (i.e., resending) past application-generated events in two cases:

(i) Synchronizing states. Microservice applications evolve over time. 
As new microservices are incorporated into the topology, it may be the case that these are required to synchronize with the current application state, as described by a developer~\cite{quote_56851950}:

\begin{displayquote} 
"When I add \textbf{new service} to a set of already running services, I need upstream dependencies to send all the messages from the past to the new service so it could \textbf{align its state} with the one of the whole system."
\end{displayquote}

(ii) Troubleshooting and Auditing.
As the communication abstraction in event-driven microservices, it is natural that the re-execution of the event flow may assist in further understanding the application behavior, as explained by the following developers:

\begin{displayquote}
(DEV \#1) I want to \textbf{replay events} on an invoice where the I want \textbf{to see all actions done} by a specific employee on the balance.~\cite{quote_49985156}
\end{displayquote}

\begin{displayquote} 
(DEV \#2) The intent for this pattern [(event sourcing)] is to provide an \textbf{audit trail of all events} that took place while the patient was in the hospital.~\cite{quote_62142695}
\end{displayquote}

\noindent\textbf{FR5. Query processing.} Developers tend to report the implementation of query processing operators at the application level. In this case, they implement queries that operate over the payload of subscribed events, as exemplified by the following SO question~\cite{quote_57328269}:

\begin{displayquote}
"I want to create a third microservice that is responsible to \textbf{join the data of ProductService and StoreService} in order to retrieve all the available products. Here, CQRS pattern seems like the best solution: I will create a materialized view and I will synchronize it using domains events published by the other 2 microservices."
\end{displayquote}

\noindent\textbf{FR6. Data replication.} We observe that the propagation of state updates is also an enabler of data replication. By subscribing to state updates in the form of events, consumers can maintain their own view of the state managed by other microservices without requiring to pull updates, thus avoiding the overhead entailed by synchronous calls (e.g., RPCs)~\cite{synapse}.

For instance, in the following question~\cite{quote_70509292}, the developer explains that
\begin{displayquote}
"Service A could raise an event whenever a new student is registering. Service B will consume the event and \textbf{stores student info in its [own] db}."
\end{displayquote}

Based on our analysis, it is worth noting that not all cases of propagation of state updates will decidedly lead to data replication, but all cases of data replication via events necessarily require propagation of state updates. 

\noindent\textbf{FR7. Cache management.} In a similar way to FR6, another reported scenario lies on maintaining caches based on subscribed events, as explained by a developer as follows~\cite{quote_57222357}.

\begin{displayquote}
"When C starts up it needs to \textbf{load all of the current data from P into its cache}, and then subscribe to change notifications. (In other words, we want to synchronize data between the services.)"
\end{displayquote}



\noindent\textbf{FR8. Task Scheduling.} We also found cases about events being used as abstractions to manage the life cycle of long-running jobs. As example scenarios, we highlight the following:


\begin{displayquote}
"That job info is placed in a RabbitMQ message and sent off by the RabbitMQ Producer [..]
A RabbitMQ Consumer receives message with the job info and calls the class that is responsible for Executing the \textbf{long running job}, the job status is updated to IN-PROGRESS"~\cite{quote_56923600}
\end{displayquote}

\begin{displayquote}
"The idea is to have the REST API immediately post a message to a queue, with a background worker role picking up the message from the queue and \textbf{spinning up multiple backend tasks}, also using queues. REST API [..] generates a GUID and attach that as an attribute on the message being added to the queue"~\cite{quote_33009721}
\end{displayquote}

\vspace{1ex}
\begin{mdframed}
\textbf{Finding 3:} Event management is applied to fulfill varied functional requirements, such as communicating data updates, composing microservices functionalities, and processing queries, indicating the heterogeneous applicability of events.
\end{mdframed}
\vspace{1ex}


\section{Event Management Challenges (RQ2)}
\label{sec:challenges}



\begin{figure*}
  \centering
  \includegraphics[width=0.9\textwidth]{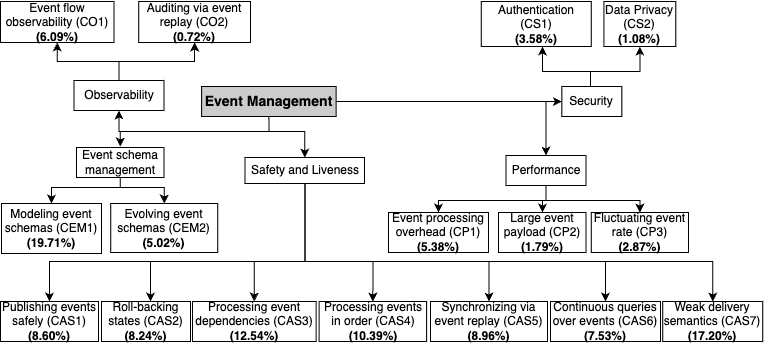}
  \caption{Overview of challenges in managing events in microservice architectures}
  \label{fig:pic_challenges}
\end{figure*}


Having discussed the state of practice of event management, we focus on the challenges developers face when managing events in their microservice deployments. Figure~\ref{fig:pic_challenges} summarizes the challenges we extracted from SO. The leaf nodes represent the specific challenges, whereas their parent nodes represent the categories to which a challenge belongs. For example, the \textit{Performance} (\ref{subsec:performance}) category comprises three specific challenges: event processing overhead (CP1), large event payload (CP2), and fluctuating event rate (CP3). In total, our analysis led to five categories and sixteen specific challenges, indicating the heterogeneity of the problems microservice developers encounter when managing events. Next, we discuss and exemplify each specific challenge by their categories. Along the discussion, it is worth noting that we refer to messaging technology as any system that enables microservices to exchange events through publishing and subscription mechanisms. 

\subsection{Safety and Liveness}
\label{subsec:safety}


Safety and liveness are properties inherent to distributed systems~\cite{lamport1977proving}. 
While safety properties prescribe that nothing bad happens, liveness properties prescribe that something good eventually happens. For instance, a traditional example of safety property is found in the 2-Phase Commit (2PC) protocol~\cite{2phase}. In 2PC, the effects of the operations of a transaction that cuts across nodes are only made available to subsequent transactions if all participating nodes agree. A common liveness guarantee example is eventual consistency~\cite{base}. In this consistency model, nodes eventually converge to the same outcomes (e.g., data item versions). When this is achieved, it is often unbounded, making it a weak consistency model.

In this section, we use these two properties to discuss related problems in the context of event management in microservice architectures. We reveal microservice practitioners' expectations when managing events, particularly when publishing, consuming, processing, reverting the effects of, and replaying events.

\subsubsection{Publishing events safely (CAS1)}
\textbf{Safety.}

The functional requirements investigated in Section~\ref{subsec:nfr} highlight that generating events is the cornerstone of event-based microservice architectures. 
Events are often raised in response to local operations, typifying a causality relation. Therefore, failures in local operations must automatically withdraw the existence of a resulting event.
However, we find that event's publishing semantics are not always crystal clear for microservice developers. We highlight as follows the most common types of issues using developers' own quotes: 


(i) Developers seek to acknowledge whether the event published has reached out to the message broker, for instance~\cite{quote_42140285}: 
\begin{displayquote}
(DEV\#1) "This operation [(publishing the event)] is asynchronous and \textbf{ALWAYS returns true}, no matter if the broker is down. How can I know that the message has reached the broker?"
\end{displayquote}

(ii) Developers wonder whether publishing an event as part of a transaction is possible, for instance~\cite{quote_62223553}:
\begin{displayquote}
(DEV\#2) "[...] does actually spring-kafka support JTA transactions and would it be enough to wrap RDBMS and Kafka Producer into \textbf{\texttt{@Transactional} [(ORM annotation)] methods}?"
\end{displayquote}

(iii) In other cases, developers wonder how to ensure atomicity semantics when publishing more than an event as part of an operation, like the user below~\cite{quote_57784098}:

\begin{displayquote}
(DEV\#3) "Let's say a command needs to append an event to both, the public and private user stream. \textbf{How can you make sure that both events have been appended?} Does the event store publish both, \texttt{SomeUserEventHapppendPrivate} and \texttt{SomeUsererEventHappendPublic}, to the event bus?"
\end{displayquote}

(iv) Developers look for alternative ways to publish events out of the critical path of the application, such as asynchronously detecting database updates and publishing these as events, like the case below~\cite{quote_39305118}:

\begin{displayquote}
(DEV\#4) "Should I have some kind of \textbf{background timed process} which will scan events table and publish events to SQS? Can this be process within WebApi application (preferable), or \textbf{should this be a separate a process}?"
\end{displayquote}

Challenges related to ensuring events are correctly published appear in 13.79\% of \texttt{Safety\&Liveness} challenges and reflect 8.60\% of the total, indicating the representativeness of the problem. 

\noindent\textbf{Discussion:} We observe practitioners seem unfamiliar with the guarantees provided by messaging technologies. As a result, they may fail to foresee the possible issues associated with ensuring consistency between two systems, in their case, the (producer) microservice and the message broker.


Recent Change-Data-Capture (CDC) tools like Debezium~\footnote{https://debezium.io} soften this challenge by dedicating a system to capture updates in a microservice private state (e.g., relational tables), offloading developers from explicitly handling the publishing of events. Each update is then eventually delivered to consumer microservices. However, apart of the need to manage an additional system, when or whether this delivery is performed is unknown, which can lead to an additional challenge as we see in Section~\ref{subsubsec:weak}.

Furthermore, some database systems like Oracle~\footnote{https://www.oracle.com} and PostgreSQL~\footnote{https://www.postgresql.org}, offer mechanisms to publish events atomically as part of transactions. However, the benefits of state encapsulation  would be jeopardized given all consumer microservices would depend on specific producer-managed databases, thus leaking the private microservice state abstraction.




\subsubsection{Roll-backing states (CAS2)}
\label{subsubsec:rollback}
\textbf{Safety.}

Once an application-generated event is published (i.e., reach out to the message broker successfully), the event becomes available for consumption by other microservices, which in turn can consume and start processing the event. However, events may be consumed and processed by multiple microservices independently, which makes it challenging to track down the possible errors across multiple microservices and react accordingly. As a result, we observe that developers express uncertainties on how to appropriately deal with failures, such as the following cases: 

\begin{displayquote}
(DEV \#1) "[...] the state modification is actually a complex operation across multiple microservices using the saga pattern, which needs to be rolled back if something fails. [...] \textbf{How to cleanup the state if the service modified it, but failed in the end}, e.g. due to system shutdown?"~\cite{quote_73928296}
\end{displayquote}

\begin{displayquote}
(DEV \#2) "When my Transaction is committed, normally I will dispatch \texttt{IntegrationEvent} (e.g. to the queue), but there is possibility that this \textbf{queue is down} as well, so previously just-committed \textbf{transaction has to be "reverted"}. How?"~\cite{quote_62364508}
\end{displayquote}


In this sense, many answers received for related questions suggest the use of compensating actions triggered via additional events; however, some developers acknowledge that these inherit the same properties of the originating problem, as summarized by the following quote: 

\begin{displayquote}
(DEV\#3) "If 3.a ([an operation triggered via event]) fails[,] a \textbf{compensation action} is performed; but \textbf{what about if it fails}?"~\cite{quote_73108942}
\end{displayquote}

We observe this challenge is significant in the context of \texttt{Safety\&Liveness} challenges (13.22\%), spanning across 8.24\% of the total questions.

\noindent\textbf{Discussion:} When the event is published, the event producer loses control over downstream processing due to the independent failure of microservices. Similarly, consumers are unaware of which microservice produced the event stream to which they are subscribed. In this case, we found some developers design specific event streams to communicate failures in the workflow, as we discuss in Section~\ref{subsubsec:rollback}.

However, publishing events that semantically represent the occurrence of a failure can also fail. That can lead "dirty" intermediate states (i.e., the effects of canceled operations) to remain exposed indefinitely. On the other hand, even though upstream microservices acknowledge a downstream failure, once the local operations are committed as part of a transaction in the database, it is often not possible to rollback the transaction. Thus, upon failure, it is unclear for developers how to properly revert those changes. 


\subsubsection{Processing event dependencies (CAS3)}
\textbf{Liveness.}

Events often carry a schema, or an event type, and are published in a stream (aka topic or queue) on which all events published must adhere to such schema. We observe an interesting trend in which practitioners attempt to match distinct but correlated events at event processing time. 
The trend is exemplified by a developer~\cite{quote_51421205}:

\begin{displayquote}
(DEV\#1) It is very straightforward to implement a [(event)] listener when \textbf{an action depends on one single event} ([stream]). [...] The problem arises when \textit{OrderService} \textbf{has to wait for more than one event} [(from different streams)]
\end{displayquote}

The developer contextualizes the case and expresses the problem then~\cite{quote_51421205}:

\begin{displayquote}
(DEV\#1) I am pooling both queues\\ \texttt{CREDIT\_AVAILABLE\_QUEUE} and\\ \texttt{INVENTORY\_AVAILABLE\_QUEUE}, and \textbf{both events has to be present} so I can finish an order. How can I coordinate so that \textit{OrderService} sees both events as only one?
\end{displayquote}

However, these dependencies across distinct events may require ordering guarantees from the messaging technology, which may not be present, as noted by another developer~\cite{quote_53270770}: 

\begin{displayquote}
(DEV\#2) If \texttt{UserCreated} \textbf{comes after} \\\texttt{ProductAddedToCart} the normal flow requires to \textbf{throw an exception} because the \textbf{user doesn't exist yet}. [...] So, the ordering problems are:
\end{displayquote}
\begin{displayquote}
No order guaranteed across events from the same event stream.\newline
No order guaranteed across events from the same ES [(event store)].\newline
No order guaranteed across events from different ES (different services).
\end{displayquote}

These ordering problem can ultimately lead to concurrency bugs in the application code, as noted by the former developer~\cite{quote_51421205}:

\begin{displayquote}
(DEV\#1) There is a minimal chance of receiving both events at the same time[,] generating \textbf{race conditions}.
\end{displayquote}



We observe that this challenge appears substantially, spanning across 20.11\& of \texttt{Safety\&Liveness} challenges. Besides, it spans 12.54\% of the total questions, highlighting the practical significance of this event processing pattern.


 
\noindent\textbf{Discussion:} Developers express some events are semantically related and thus cannot be processed independently. However, developers fall prey to weak ordering semantics across event streams. Another challenging factor is that the delivery of events can experience arbitrary delays due to network partitions, overload of computational resources, network jitter, and even the failure of producer microservices. These aspects together pose a challenge in enforcing event dependencies across distinct event streams.

This event processing characteristic may find resemblance with the so-called workflow patterns, posited for business processes and web service compositions~\cite{workflow_patterns} a few decades ago. Recent microservice frameworks, like Dapr~\footnote{https://docs.dapr.io}, has included in their API support for some workflow patterns, such as fan-out/fan-in~\cite{fanoutfanin}. However, workflow patterns often require either a domain-specific language or a common framework where all microservices are implemented with, as well as the explicit specification of workflows, including event producers and consumers. These contrasts with the polyglot nature of microservices and the decoupling through events seek by microservice developers. Thus, it is an open question whether and how workflow patterns can mitigate the challenges of processing event dependencies. 

\subsubsection{Processing events in order (CAS4)}
\textbf{Safety.}

While analyzing the questions, we noticed that practitioners often expect events to be processed in order for every consumer. However, in many cases, this expectation is not met. Unlike the previous challenge, this challenge affects single event streams (i.e., all conforming to the same event type) rather than distinct events. A developer exemplifies the problem as follows.


\begin{displayquote}
(DEV\#1) Suppose a product is ordered and it is \textit{id} 80, and a series of sequential update events fired from product service to order services for that particular product [...]
\textbf{The end price should be 100 for that product, but sometimes these events are processed in random order}~\cite{quote_67612615}
\end{displayquote}

The developer expects that the product price observed by the \textit{Order} microservice must eventually be in sync with the product's state in the producer (i.e., \textit{Product} microservice). Therefore, any processing order not leading to a price of $100$ as the final state is deemed incorrect. In some cases, practitioners acknowledge that an in-order processing guarantee cannot be enforced in consumers due to the uncertain semantics of producers:


\begin{displayquote}
(DEV\#2) Since each microservice has its own event table and asynchronous worker, \textbf{we cannot guarantee that events will be sent in the sequence} in which the corresponding state changes occurred in their respective microservices.~\cite{quote_47516458}
\end{displayquote}

This challenges appears in 16.67\% of \texttt{Safety\&Liveness} challenges and accounts for 10.39\% of the total challenges observed, highlighting the difficulty of ensuring events are processed in the expected order.

\noindent\textbf{Discussion:} Although the majority of messaging systems guarantee the delivery of events from the same stream in order, such as Kafka through the concept of a topic partition~\cite{kafka_partition}, that does not exclude producers and consumers from still processing the events accounting for the expected processing order semantic. For example, for producers consisting of concurrent threads writing to the same partition, even though Kafka serializes the messages, the correct order that the consumer must process the messages is not well defined. Similarly, suppose multiple consumer threads pull messages concurrently from Kafka. In that case, if the order of messages across different pulls matters, it is complex and error-prone for general developers to reassemble the original stream order at the consumer side.


In most questions analyzed with this particular challenge, we could not identify whether the message broker delivers the messages out of order or the microservice processes the event arbitrarily (not accounting for the delivery order). As the quote below exemplifies~\cite{quote_52632129}, event processing semantics appear unclear to some microservice developers.

\begin{displayquote}
(DEV\#3) In [(omitted technology,)] \textbf{messages are best-effort ordering, still no idea of what they mean}. [...]
Does it means that[,] giving n copies of a message[,] the first copy is delivered in order[,] while the others are delivered unordered compared to the other messages' copies? Or "more that one" could be "all"? 
\end{displayquote}

\subsubsection{Synchronizing states via event replay (CAS5)}
\textbf{Safety.}

As described in Section~\ref{subsec:nfr}, the need for replaying events often arises due to new microservices being introduced in the system. In this context, a developer explains an interesting challenge~\cite{quote_71160328}:


\begin{displayquote}
(DEV\#1) Service "A" creates event message to inform other services about changes [...]
\textbf{Newly introduced Service "D" also needs to replicate data} coming from service "A". Service "D" needs all historic data[.]
\end{displayquote}

The developer then continues expressing the impedance found.

\begin{displayquote}
([...]) other services were already running for a while, and \textbf{service "A" only broadcasts new changes}. What would be the correct solution to populate newly added service with historic data?
\end{displayquote}

Furthermore, as microservices supposedly fail independently, it may be necessary to replay events from the point in time the subscriber microservice(s) failed, as described by another developer~\cite{quote_65425071}:

\begin{displayquote}
(DEV\#2) I'm wondering what should happen if one of these events can not be delivered due to an error [...] On republishing events, should all messages be republished to all topics or would \textbf{it be possible to only republish a subset}?
\end{displayquote}


Challenges involving replaying events also appear significantly in \texttt{Safety\&Liveness} challenges (14.37\%), accounting for 8.96\% of the total challenges, suggesting mechanisms for safely replaying events are strongly seek by developers but missing in practice.

\noindent\textbf{Discussion:} Along the analysis, we find that the need for replaying events arises not only from software evolution, including adding new microservices and migrating to a microservice architecture but also when recovering from failures and fixing the outcome of bugs. However, developers express uncertainties about how to proceed properly in the cases above and end up implementing several ad-hoc mechanisms at the application layer to fulfill these. As a result, these mechanisms often result in rework and create additional issues in the application.


\subsubsection{Continuous queries over events (CAS6)}
\textbf{Liveness.}

We observe that practitioners commonly carry out continuous queries consisting of various query operators (e.g., filter, join, and aggregations)~\cite{query} over the payload of incoming events to extract valuable information. However, the nondeterminism of asynchronous events (i.e., the arrival of events may experience large delays or even never arrive) poses challenges in ensuring correct query processing results, as described by a developer~\cite{quote_57328269}:

\begin{displayquote}
(DEV\#1) When I update a \textit{Product} entity or \textit{Store} entity there is no problem to sync the view because I already have data on it. But what about when I receive a \texttt{ProductCreated} event? 
\end{displayquote}
\begin{displayquote}
This event has only information about product and nothing else, so \texttt{stock}, \texttt{store\_name}, \texttt{store\_address} \textbf{will be NULL}. 
\end{displayquote}

\begin{displayquote}
How can I save this event in my view? \textbf{Should I save incomplete data somewhere else} and update my view \textbf{when I will receive complete data}?
\end{displayquote}

Furthermore, as software evolves over time, changing requirements can also undermine providing correct query results based on events, as exposed by another developer~\cite{quote_67074772}:

\begin{displayquote}
(DEV\#2) I have \textbf{new requirement} - calculating maximum all time temperature per sensor. 
\end{displayquote}

\begin{displayquote}
I have prepared \textbf{new microservice} that creates \texttt{KTable} ([Kafka table abstraction]) aggregating temperature (with max) grouped per sensor. 
\end{displayquote}

\begin{displayquote}[67074772]
Simply deploying this microservice would be enough if input topic had infinite retention, \textbf{but now maximum would be not all-time}, as is our requirement. [...] How to design the solution?
\end{displayquote}

The impediment found above relates to the inner characteristic of stream processing engines, that necessarily rely on windows and notions of time to avoid querying all the historical data in order to provide fast responses~\cite{hueske2019stream}. Continuous queries over events appear frequently in Safety\&Liveness challenges (12.07\%) and corresponds to 7.53\% of the total questions, highlighting the significance of this challenge.

\noindent\textbf{Discussion:} For over two decades, continuous queries and stream processing systems have been a mainstream research topic in the data management community~\cite{fragkoulis2023surveyevolutionstreamprocessing}, resulting in industry-strength solutions like Flink~\footnote{https://flink.apache.org} and Kafka Streams~\footnote{https://kafka.apache.org/documentation/streams}. These systems allow users to declare a query that is continuously updated based on incoming streams. However, we notice microservice developers often do not mention stream processing systems and prefer patterns like CQRS, not realizing that providing correct query results with this pattern incurs additional challenges. Besides implementing the query operators, developers must ensure that duplicate events (we further discuss in Section~\ref{subsubsec:weak}), crashes, and delays do not impact the query results.

Stream processing systems are designed to handle the gamut of problems that arise in maintaining continuous queries. We conjecture microservice developers are either unaware of or may find difficult to map their application logic to stream processing operators.

\subsubsection{Weak delivery semantics (CAS7)}
\label{subsubsec:weak}
\textbf{Liveness.}






Messaging technologies often guarantee at-least-once delivery by default, leaving the responsibility of achieving at-most-once or exactly-once delivery semantics to application developers~\cite{fragkoulis2023surveyevolutionstreamprocessing}. However, by resorting to at-least-once delivery semantics, developers can face unpredictable challenges, as exemplified by the developer as follows~\cite{quote_63391504}:


\begin{displayquote}
(DEV\#1) I was able to collect the data [(published as events)] from another microservice, but I noticed that if I terminate the [(consumer)] process and run it again, I get the data back, PLUS another copy of the same data right under it. [...] \textbf{I got the payload sent to me twice, and I only want to see it once}.
\end{displayquote}

As the developer's quote shows, a problematic aspect of at-least-once delivery is that events can be delivered more than once. That forces consumers to either make their 
microservices
idempotent or fall prey to the undesired effects of duplicate event processing. The following microservice developer expresses another interesting case~\cite{quote_42230797}: 


\begin{displayquote}
(DEV\#2) The only way I have \textbf{to know} that the \textbf{payment has been processed} by \textit{Payment} service \textbf{is by expecting an answer event} (\texttt{Payment ok|failure}).
\end{displayquote}

However, in the absence of an event that confirms that a payment has been processed, the developer comes up with its own customized solution:

\begin{displayquote}
If it hasn't gotten an answer in some time, [that] \textbf{forces me to implement a retry mechanism} in the \textit{Order} server, [that is,] retry with a new \texttt{Payment} event.
\end{displayquote}

The developer ended up dealing with an unexpected outcome:

\begin{displayquote}
\.[However,] this also \textbf{forces me to take care of duplicated messages} in \textit{Payment} service in case they were actually processed but the answer didn't get to the \textit{Order} service.
\end{displayquote}

Another popular issue on resorting to events for asynchronous processing lies on the fact that computations are often initiated based on client requests. These online requests require timely response to users and challenges developers to match the eventual arrival of asynchronous events (i.e., the response of the client-triggered asynchronous computation) with the corresponding waiting client. A developer describes the problem below~\cite{quote_54451013}:

\begin{displayquote}
After publishing the event[,] my booking \textbf{service can't block the call and goes back to the client} (front end). 
\end{displayquote}

\begin{displayquote}
How does my client app will have to check the status of transaction? Does it poll every couple of seconds? Since this is distributed transaction and \textbf{any service can go down and won't be able to acknowledge back}. 
\end{displayquote}

\begin{displayquote}
In that case how do my client (front end) would know since \textbf{it will keep on waiting}.
\end{displayquote}

Dealing with weak delivery semantics is the most popular challenge in \texttt{Safety\&Liveness} category (27.59\%). By appearing in 17.20\% of total the questions, that evidences the difficulty of managing events under weak delivery semantics.

\noindent\textbf{Discussion:} The loss of events may be caused by many issues. Besides consumers crashing, during a rebalance in Kafka brokers, consumers can experience delays in event delivery~\cite{kafka_issues}. Besides, the network can become unstable, leading to delays in network package deliveries. 
In an attempt to ensure the completeness of operations across microservices, developers often rely on custom-made, ad-hoc solutions, such as the retry mechanism shown above. However, retries can lead to the additional burden of dealing with duplicate events, which forces developers to implement additional mechanisms at the application layer to impede duplicate events from being processed, only exacerbating the problem.

\vspace{1ex}
\begin{mdframed}
\textbf{Finding 4:} Developers encounter a myriad of challenges on ensuring events are processed correctly. These span the entire life-cycle of an event in a microservice architecture, including queuing, delivering, storing, processing, and synchronizing events. 
\end{mdframed}
\vspace{1ex}

\subsubsection{Implications}

Given the wide range of safety and liveness challenges, we discuss the implications in two groups. The first covers event publishing and delivery semantics and the second relates to the challenges of synchronizing microservices' events. 

\noindent\textbf{CAS 1, 2, 4, and 7.} Developers lack a comprehensive specification to guide them with the variety of event processing semantics and in dealing with failures in asynchronous, distributed microservices, suggesting they are in need for supporting tools. For messaging technology providers and framework developers, their documentation can emphasize the publishing and delivery semantics offered more clearly, including, but not limited to, the possible failure scenario microservice developers must consider and example workarounds to mitigate some of their effects in the microservice state. In addition, these should be provided in a language suitable for developers who are not familiar with distributed and asynchronous systems.

For database providers, explicit APIs for rolling back microservices' underlying databases to a state prior to processing a given event can help alleviate the burden of dealing with online failures in event-based workflows. Researchers can systematically characterize the guarantees provided by popular messaging technologies and enhance code analysis tools to better alert developers about potential hidden shortcomings of the different guarantees.

\noindent\textbf{CAS 3, 5, and 6.} The need to process and match distinct event streams generated by different producer microservices spans different functional requirements, such as replicating data, processing distinct but correlated events, and maintaining queries over event streams. Together, they form 26.83\% of the total challenges, highlighting the practical significance of these event processing patterns. To fulfill these, developers often hold the assumption that events will arrive in a timely and exactly once manner. However, events may either take arbitrary time to arrive or never arrive. Besides, even if events arrive, they may be repeated. The mismatch between expectations and reality leads to frustrations in implementing the requirements above. Thus, microservice developers must be vigilant with designs that favor the existence of such event processing patterns. Rethinking service boundaries~\cite{10160171} may lead to some of the events involved in intricate event processing patterns being merged or eliminated.

Providers of messaging technologies and cloud computing services, and framework developers can offer further guidance at the documentation and API levels to alert developers regarding the potential dangers of some event processing patterns. Potentially, message technologies can devise better APIs to support developers making sure event dependencies are met. For example, an API to wait for multiple dependant events instead of individual events.


Researchers can improve static and dynamic analysis tools to detect certain code properties and execution traces that may lead to the harmful scenarios discussed. 
Another direction could be developing automatic tools to enforce safety properties on event processing. For example, Lesniak et al.~\cite{stream_constraints} 
extend message queuing system with a programmable interface where developers can specify event processing order which will be enforced automatically.

\noindent\textbf{All CAS.} Researchers can devise or extend existing programming models to account for use cases that necessitate correlation between distinct event streams and abstract publishing and delivery semantics from the microservice code. For instance, they can take inspiration from the dataflow model~\cite{dataflow}, commonly used in stream processing engines, to accommodate the missing dynamic topology, non-blocking, isolated failure model found in real-world microservices. That can simplify the programmability of microservices and favor the adoption of tricky event stream processing patterns. Additionally, this programming model can unify the design and deployment of event-based microservices across cloud providers and execution platforms to facilitate the migration and hybrid execution of microservice applications across message technology providers, an emerging trend in cloud computing~\cite{google_cloud}.

\subsection{Event Schema Management}

Microservice practitioners model events for external interaction. In other words, the decoupling nature of a microservice design requires producers to encode all the necessary data so that consumers do not need to query the producer for additional data when processing such an event. Furthermore, the events generated by microservices must comply with a data type to be correctly processed by consumers, and, as with any other data type, it can evolve over time. This section discusses the challenges associated with modeling and evolving event schemas in microservice architectures.

% ...



\subsubsection{Modeling event schemas (CEM1)}

We observe that how events are modeled and what they represent semantically may dictate the design and performance of microservices. For example, a developer looks for recommendations for event design explaining the following scenario~\cite{quote_45486658}:

\begin{displayquote}
Say you have a micro-service architecture where multiple services produce and consume \textit{unit\_statuses}. \textbf{There are multiple ways to design this}.
\end{displayquote}

The person continues explaining some options in mind:
\begin{displayquote}
1. Create a \textbf{generic topic \texttt{unit-status}} and make services consume and produce messages on this topic.  [...] \newline
2. Create a \textbf{specific topic for each status}, for example \texttt{unit-status-created}, \\\texttt{unit-status-packaged}, \texttt{unit-status-loaded}, \texttt{unit-status-deleted}, etc. 
\end{displayquote}

In sequence, the practitioner explains the drawbacks of each:

\begin{displayquote}
1. This has the consequence that \textbf{you consume your own messages and have to filter them}.
2. Requires a \textbf{code or configuration change} in potentially all service when a new status topic is added.
\end{displayquote}

In another question~\cite{quote_53025888}, a similar inquiry is presented where a practitioner wonders about two event design approaches:

\begin{displayquote}
(1) \textit{A} sends a message which contains event and related entity id like: \texttt{entityCreated} \{ \texttt{entityID}: 1234 \}. \textit{B} consumes this message and \textbf{if it needs further information, it fetches this from A} with \texttt{entityID} \newline
(2) The message not only contains the information above, but \textbf{also metadata} like: \texttt{entityCreated} \{ \texttt{entityID}: 1234, \texttt{SomeFieldKey}: someFieldValue, ... \}
\end{displayquote}

The practitioner continues addressing the pros and cons of each and wonders which one to choose:
\begin{displayquote}
(1) Pros: \textbf{Less network usage; Always the same structure of messages}. Cons: If information from A is needed on demand there \textbf{must be some mechanism to catch}, e.g. network failures. \newline
(2) Pro: \textbf{Information is already there}. Con: What if the \textbf{attached information is not enough}?
\end{displayquote}

Although practitioners seem aware of some of the consequences different event modeling options bring to the application design, the excessive number of questions, 19.71\% of the total, suggests developers encounter uncertainties in deciding on an optimal event design.

\noindent\textbf{Discussion:} Differently from relational model~\cite{codd}, which provides a formal foundation that allows database designers to reason about the possible performance implications of a data model (e.g., the level of normalization dictates whether additional \texttt{JOIN} operations are necessary to retrieve certain information from the database), event design is still an open problem. As a result, developers navigate through many possible event designs that, if not carefully thought out, may impact the design of the application and the performance of event-triggered operations.


\subsubsection{Evolving event schemas (CEM2)}

We also observe some microservice practitioners share uncertainties on how to deal with the possible effects of event schema evolution properly. In particular, as events are often parsed into microservices' own data models, practitioners express concerns over the impact of event schema changes in their microservice states. For instance, upon a new attribute being incorporated into an event's schema (user's \texttt{address}), a developer wonders how to update a microservice state retroactively~\cite{quote_59923240}:

\begin{displayquote}
(DEV\#1) I can update my event and \textbf{now include the \texttt{address}}, but it will only work for new users, \textbf{the old ones will have \textit{null} addresses}. Should I scan the whole database and manually dispatch an event for each user?
\end{displayquote}

On the other hand, although events are considered abstractions to decouple microservices, some practitioners report event schema evolution impacting dependent microservices~\cite{quote_68781332}:

\begin{displayquote}
(DEV\#2) services are tightly coupled by [event] schema and \textbf{can causes errors and this contradicts with event sourcing goal}. How can we address this problem?
\end{displayquote}

Evolving event schemas also appear as an important challenge, corresponding to 20.59\% of the questions about managing event schemas in microservices and accounts for 5.02\% of the total questions.

\noindent\textbf{Discussion:} We observe that developers are aware of the risks associated with schema changes and their potential impact on microservices dependent on the modified event type. If the type of incoming events does not match the expected event type in a consumer microservice, the processing of this event processing may fail. However, developers have not found a systematic solution to either minimize the impact of or comprehensively manage the evolution of event schemas across the network of dependent microservices.

State-of-the-practice data serialization formats like Apache Avro~\footnote{https://avro.apache.org} can potentially ease this challenge. However, such tools require a global adoption across all microservices, forcing every event producer and consumer to employ schema and data contract enforcement. Besides, adopting such tools is not always possible because the messaging technology adopted may not provide native support. In overall, this remains an open challenge in practice.

\vspace{1ex}
\begin{mdframed}
\textbf{Finding 5:} Developers find no principled ways to model events accounting for performance and decoupling trade-offs. Systematic support for managing event schema changes is also an open problem.
\end{mdframed}
\vspace{1ex}

\subsubsection{Implications} With the lack of automatic support for event schema evolution, microservice developers can isolate failures caused by event schema mismatches and log them appropriately for later reconciliation. Messaging technology providers can supplement their documentation with more technical examples of how developers can isolate the microservices' application code from the impact of event schema changes. Further guidelines on appropriately rolling out event schema changes without impacting individual microservices in the context of their technologies can also benefit developers~\cite{OVEREEM2021110970}

Researchers can develop analysis tools that holistically map the microservice event topology, matching producer and consumer microservices through the events exchanged. The tool can incorporate metrics to aid developers in reasoning about the trade-offs of different event designs. Furthermore, researchers can provide a formal foundation for modeling events in microservice applications. The model must allow for reasoning about the performance trade-offs of including specific attributes in events. Lastly, automatic support for event schema evolution across microservices is an open and complex problem that researchers could tackle.

\subsection{Performance}
\label{subsec:performance}

Performance is systematically cited as an important factor for adopting microservice architectures~\cite{base,8705256,LaignerKLSO20}.
Developers seek to reap the benefits of having the application decomposed into independently scalable building blocks~\cite{vldb}. In an architecture based on events, though, the performance of individual microservices is tightly coupled with its capacity to process and dispatch application-generated events at a rate that does not compromise the expected performance of the downstream microservices in the event flow. As a result,
disruptions in the event flow can lead to critical performance impacts. In this section, we analyze such cases from the lens of microservice developers.



\subsubsection{Event-processing overhead (CP1)} 

Developers report particular event processing scenarios that ultimately lead to a performance penalty. For example, in connection with \textbf{CAS5}, some practitioners report the need to replay events to fulfill a given requirement, but that ends up introducing additional load to the application~\cite{quote_53949531}:





\begin{displayquote}
The validation of a \textbf{reservation request needs to know the previous and following reservations} (within 3 hours from the booking time of the incoming request)
\end{displayquote}

\begin{displayquote}
\textbf{if I ask for a reservation for tomorrow, the system will replay reservations from 6 months ago} that usually are not related with the incoming request 
\end{displayquote}

The developer then highlight the overhead:
\begin{displayquote}
This leads to inefficiencies over time as result of the \textbf{huge amount of unnecessary events that are replayed}. I thought to solve it using daily snapshots but it seems the wrong way to do it.
\end{displayquote}

In a similar way, in connection with CAS3, other developers acknowledge there could be a performance penalty on leveraging events for replicating data across microservices. For example~\cite{quote_42528718}:

\begin{displayquote}
The \textit{Invoices} microservice must listen to all \\\texttt{ContactCreated} and \texttt{ContactDeleted} events in order to know if the given recipient id is valid.
\end{displayquote}

\begin{displayquote}
Then \textbf{I'd have thousands of Contacts} within the \textit{Invoices} microservice, \textbf{even if I know that only a few of them will ever receive an \texttt{Invoice}.} Is there any best practice to handle those scenarios?
\end{displayquote}





Practitioners express uncertainties in dealing with the overhead of processing events and their possible performance impacts. This challenge is the most prominent in  the \texttt{Performance} category (53.57\%), and spans 5.38\% of the total questions. This suggests that microservices may execute with suboptimal performance in real-world deployments.

\noindent\textbf{Discussion:} At first sight, the overhead of the first case can be potentially mitigated by storing the events processed in the microservice's private state. However, this technique introduces hidden dangers. For example, keeping track of the last stored event is nontrivial and can ultimately lead to losing an event due to crashes. Besides, developers tend to favor the data shipping paradigm~\cite{bykov2010orleans}. By offloading state management to another system (e.g., database system, message broker, or cloud object storage), applications can remain stateless, decreasing their complexity. In the second case, the overhead is triggered by the lack of systematic support for data replication through event management.

\subsubsection{Large event payload (CP2)} 


We observe developers also present concerns over the size of the event payload generated and consumed by different microservices. For instance, a common concern lies in whether publishing events with large payloads into the message layer entails a good practice:



\begin{displayquote}
(DEV \#1) My question is, is it wise to stream such file over event bus from Service A to API gateway? (\textbf{File may get as large as 100 MB})~\cite{quote_51665264}
\end{displayquote}

\begin{displayquote}
(DEV \#2) If an event needs to pass a \textbf{very large volume of data} to the next Saga event, how is this done in terms of the request structure? Is it divided into multiple Sagas for example (as a result pagination type)?~\cite{quote_48487098}
\end{displayquote}

\begin{displayquote}
(DEV \#3) I think we will need to go with second option i.e. not sending data in the event due to the \textbf{potential size of the event data}. ~\cite{quote_64049385}
\end{displayquote}

To circumvent the potential performance degradation incurred by processing large event payloads, some developers look for alternatives to the solo event-driven approach but end up encountering other issues:

\begin{displayquote}
(DEV \#3) I was also thinking about adding the URI in the event so that the service just needs to call it. Only issue I can see is that \textbf{how would the service know what type to deserialize} the response to?~\cite{quote_64049385}
\end{displayquote}

The performance impacts of generating, processing, and storing large event payloads account for 17.86\% of \texttt{Performance} challenges. This challenge represents 1.79\% of the total questions and highlights that developers find difficulties on dealing with large event payloads and their consequences to microservice performance.

\noindent\textbf{Discussion:} Message brokers and modern log processing systems are not designed to handle messages with large payloads natively. Message brokers have historically targeted systems' integration cases, which usually do not require large event payloads~\cite{base}. Log processing systems, like Kafka, for example, were initially designed for the timely processing of logs extracted from processes running in distributed servers~\cite{kreps2011kafka}. Similarly to message brokers, though, it was never a design goal to accommodate the processing of large log payloads. Native support for these types of objects is found in cloud storage services and specific column-oriented database systems, like \texttt{BLOB} type in PostgreSQL.~\footnote{https://www.postgresql.org/docs/16/largeobjects.html} Therefore, the questions analyzed in StackOverflow suggest that microservice developers may benefit from either redesigning their computations to manage large objects in appropriate storage systems or rethinking the granularity of their event payloads when they grow arbitrarily. This may involve breaking down application components into smaller, finer-grained tasks to decrease the size of events transmitted across microservices.


\subsubsection{Fluctuating event rate (CP3)}

Developers describe challenges associated with dynamic workloads and their impact on the observed event processing rate.

\begin{displayquote}
(DEV \#1) In case of \textbf{pressure over the system}, can service B communicate to service A to \textbf{slow down}, and A will react to this by \textbf{not accepting more requests} from clients, till B decides it can continue? Is this something that can be achievable using [(omitted library)]?~\cite{quote_57181251}
\end{displayquote}

\begin{displayquote}
(DEV \#2) I want to \textbf{restrict the users from publishing a new message} on this topic to \textbf{prevent} my system from \textbf{choking after a certain limit}.
\end{displayquote}

\begin{displayquote}
For example, if the number of unacknowledged messages in the topic is already \textbf{more than or equal to 10000}, then I want to give a bad input \textbf{exception} or something to restrict users from flooding my queue.~\cite{quote_74064658}
\end{displayquote}


\begin{displayquote}
(DEV \#3) \textbf{Traffic is increasing} and we've noticed that events \textbf{spend a lot of time in queue}. We need to \textbf{process events faster}. [...] Is it possible to implement circuit breaker or a failover mechanism for an async call?~\cite{quote_68307199}
\end{displayquote}


The challenges about adapting microservices to handle increased influx of events represent 28.57\% of \texttt{Performance} questions. Representing 2.87\% of the total of questions, this additional performance challenge strengthens the perception that microservice developers encounter difficulties on reaching performance goals in event-driven microservice architectures.

\noindent\textbf{Discussion:} With traditional synchronous communication paradigms, like RPC and HTTP requests, an application can usually define a threshold of the maximum allowed concurrent connections. In event-based architectures, though, there is an indirection: the events are often stored in the event management layer first and only later forwarded to (or pulled by, depending on the system) consumers. This decoupling introduces challenges for consumer microservices to communicate whether producers must decrease the event generation rate or even refrain from generating new events. On the other hand, modern event management systems, like Kafka, provide abstractions to parallelize the consumption of events through partitions. However, developers must configure the unit of parallelism according to their requirements, which may not be trivial for newcomers. 


\vspace{1ex}
\begin{mdframed}
\textbf{Finding 6:} Developers encounter challenges on managing the event rate across producers and consumers. This only exacerbates in the presence of event replays and large event payload sizes.
\end{mdframed}
\vspace{1ex}

\subsubsection{Implications}
Message technology vendors can provide efficient and proactive event-processing techniques and algorithms that mitigate or decrease the impact of fast producers and slow consumers. As mentioned in the discussion of individual challenges, developers may benefit from rethinking the granularity of their microservices and the content included in their event payloads to escape from adding overhead to their event processing pipelines. Developers may also be aware that, as events flow through microservices, the events generated by each microservice may include additional data. That suggests developers may track the event payload sizes more carefully as they flow across possibly many microservices. 

Researchers can investigate whether state-of-the-art approaches for scaling distributed systems also apply to event management in microservices. Besides, researchers can develop systematic tools to mitigate the effects of large payload sizes in event-based microservice workflows. That could involve optimal deserialization of event payloads (e.g., only deserializing the subset of attributes that the microservice requires), advanced compression tools that mitigate the overhead of growing events, and transparent removal of unused event attributes.

\subsection{Observability}
\label{subsec:observability}

Observability is a critical concern in modern software development because it allows developers to track important application behavior and metrics, such as application exceptions and memory usage, respectively~\cite{Visualizing}. In distributed systems, such as microservice architectures, observability becomes more critical since problems can possibly span across multiple components that execute independently in a decentralized manner~\cite{Visualizing,observability_microservices}.

In our analysis, we were surprised to find that microservice developers leverage events not only as an abstraction to process operations asynchronously but also to track the progress of operations across multiple microservices and troubleshoot the complex interplay of multiple microservices. In this section, we describe these patterns and their associated challenges.


\subsubsection{Event flow observability (CO1)} 

Although tools to observe synchronous requests (HTTP or RPC-based) crossing microservices enjoy apparent consolidation, developers express challenges in monitoring the asynchronous event flow across microservices, as exemplified by the two following developers' quotes:

\begin{displayquote}
(DEV \#1) What is something we can do to have an \textbf{end to end understanding} from producers to consumers, etc? Mainly for troubleshooting purposes/change management.~\cite{quote_65500926}
\end{displayquote}

\begin{displayquote}
(DEV \#2) What is the best way \textbf{to track the JSON as it flows} through many microservices down stream in an event driven way?~\cite{quote_63699141}
\end{displayquote}

Another example presents a developer wondering whether an observability tool can embrace events:
\begin{displayquote}
(DEV \#3) Will [(omitted tool)] \textbf{be able to trace requests done over the eventbus?} The tracing page says that headers need to be propagated through in http or grpc - but the eventbus sends messages via tcp -- does that mean that [(omitted tool)] will not be able to trace requests and show the visualisation tools[?]~\cite{quote_51123104}
\end{displayquote}


Furthermore, we found that issues in the message broker only exacerbate the challenge of monitoring how healthy the event flow is, as explained by a developer~\cite{quote_67565608}:


\begin{displayquote}
For any reason \textbf{the messages get stuck after some time [...] There is nothing in the log files nor the OS event log.} I have to restart the ([omitted technology]) service in order to "reanimate" it. 
\end{displayquote}

\begin{displayquote}
Afterwards[,] all stuck messages will be processed and \textbf{everything is working fine until the next "accident".} [...] Does anybody has an idea what I could check additionally to find out what the problem is?
\end{displayquote}

Observing the event flow appear significantly in \textit{Observability} challenges (89.47\%).
This highlights the difficulties developers find to understand the progress of events as they flow across microservices.

\noindent\textbf{Discussion:} Application-generated events serve as natural progress markers. It is a natural choice for developers to obtain an end-to-end overview of the application execution based on these events. However, developers encounter a twofold problem. On the one hand, developers find that state-of-the-art observability tools do not natively support tracking application-generated events. 
On the other hand, when disruptions in the event flow are not caused by crashes or bugs in the microservices themselves, developers have difficulties in finding the root problem. 
Although messaging technologies are only part of the intricate and heterogeneous components in microservice deployments~\cite{vldb}, they are a core enabler of event-based interactions in microservices. Thus, timely detection of message broker failures and their relationships with microservice disruptions are key to observability in event-based architectures.


\subsubsection{Auditing via event replay (CO2)}

In CAS5, we discussed how developers leverage events in an attempt to synchronize data across microservices. As a related pattern, we also found that 
developers use the events as an abstraction to reproduce some application behavior. Similar to CAS5, developers end up facing difficulties in ensuring the correctness of the process, as explained by a developer~\cite{quote_62142695}:

\begin{displayquote}
(DEV\#1:) If the VisitId was deleted across all services we could just \textbf{replay the events} one at a time, in order, and reproduce an exact copy of the original record.
\end{displayquote}

The developer then mentions an impedance commonly expressed by another practitioner across the questions:

\begin{displayquote}
I've been using [(omitted message technology)] for the stream itself [...] 
The issue though is that \textbf{this is not suitable for a replay store} - only delivery of the event messages.
\end{displayquote}

Another developer warns about the hidden dangers of such a method~\cite{quote_49986434}:

\begin{displayquote}
(DEV\#2:) What's maybe not a good idea is trying to \textbf{recover the current state} of your domain model by replaying an arbitrary set of your events.
\end{displayquote}
\begin{displayquote}
[..] Remember, \textbf{an event isn't usually a complete representation of the state of the model after the change}, but rather a description of the things that changed. 
\end{displayquote}

Auditing past microservice executions through events to observe and further understand their behavior appear in 10.53\% of \texttt{Observability} challenges. 
This suggests that current technologies offer insufficient abstractions to support event replay effectively.

\noindent\textbf{Discussion:} As events often represent intermediate microservice states, they appear as convenient abstractions to developers seeking to troubleshoot past application behavior. However, in the same vein as CAS5, this practice incurs hidden challenges due to the lack of appropriate support from current messaging and framework technologies. For instance, as discussed in CAS7, mismanaging the processing of duplicate events and their possible effects on the application can lead to corrupting a microservice state. Another example, based on CEM1, is
that microservices may even fail to replay events due to the divergence of event schemas.

\vspace{1ex}
\begin{mdframed}
\textbf{Finding 7:} Developers find difficulties on observing and thus reacting upon disruptions that occur in the event flow across microservices. These disruptions may not only be caused by microservices, but also message brokers and network partitions.
\end{mdframed}
\vspace{1ex}

\subsubsection{Implications}
Messaging technology providers can evolve their systems and APIs to provide native integration with industry-strength observability tools. The integration must account for the needs of developers in their real-world cases as discussed above, including but not limited to specific metrics such as event processing and acknowledgment delays from consumers.

For developers, as state-of-the-art abstractions render limited support for safely replaying events, developers can use a "staging" environment (similar to testing environments with testing, staging, and production ~\cite{staging}) to audit their event-triggered operations. However, generating an application and message broker state from a previous point in time for replaying can be challenging.

In the tool development landscape, opportunities for researchers are:
(i) They can co-design static analysis tools and dynamic analysis of event streams for a holistic monitoring of microservice event streams. In particular, this can aid developers by automatically matching correlated events and associated service disruptions in the event flow.
(ii) They can develop techniques for isolating re-execution of events from impacting the actual microservice states. Recent tools~\cite{r3} allows for replaying operations in database-backed applications. However, it is unclear how the approach can be applied to events and distributed states of microservices.

\subsection{Security}
\label{subsec:security}

Security is another critical concern in event management in microservices. In this section, we discuss developers' requirements and challenges when ensuring that event ingestion, processing, and storage meet security constraints.





\subsubsection{Authentication (CS1)}



Although events are often an abstraction that is only internally recognized by the microservices, some events may be generated by end users. That requires making public APIs available, which necessarily exposes the microservice to security vulnerabilities, as exemplified as follows:

\begin{displayquote}
(DEV\#1) Because I \textbf{don't check authorization} on websocket open, in theory this approach is vulnerable to a \textbf{dDos attack}, where an attacker simply opens as many sockets as they can.~\cite{quote_61737655}
\end{displayquote}


On the other hand, even when APIs are safeguarded with authentication methods, developers are concerned about aligning the validity of the authentication mechanism with the event flow.

\begin{displayquote}
(DEV\#2) In case of event driven application, it's possible \textbf{to ensure the token is valid?} In case of failure, the user clicks on button and an event is written but the processing of this will be hours later. How can I process the event with the user credentials?~\cite{quote_29148547}
\end{displayquote}


\begin{displayquote}
(DEV\#3) if I change my communication mechanism to using \textbf{async messaging} (e.g. RabbitMQ), how would I now authenticate the event to the scope of the user who initiated the event.~\cite{quote_62109564} 
\end{displayquote}

Challenges on ensuring event-based workflows and event-triggered computations execute securely appear substantially in \texttt{Security} challenges (76.92\%). Spanning across 3.58\% of the total questions, that suggests developers fall short on a holistic solution to secure their event-driven microservices.

\noindent\textbf{Discussion:} Aligning authorization mechanisms and event processing is non-trivial. One obstacle is that the authorization and messaging mechanisms are often provided by different systems, resulting in a system integration issue. In particular, developers working with microservices need to handle authentication token life cycles and make sure that any events related to the token are in line with its validity. This process can add complexity to the solution.

\subsubsection{Data Privacy (CS2)}

Data privacy emerged as a key requirement in modern applications to prevent leakage of user data and potential violation of data privacy laws such as the General Data Protection Regulation (GDPR)~\footnote{https://gdpr.eu/}. Data privacy laws prescribe that users can request the removal of their data at any moment, and firms are usually given a deadline for fulfilling the request. As discussed earlier, in microservices, data flows from the multiple event streams to many microservice consumers, which in turn apply operations on their private states. This heterogeneous data placement and movement can make managing data privacy challenging. As exemplified by the following quote~\cite{quote_57784098}, we observe that microservice developers express uncertainties about handling data privacy in event management.



\begin{displayquote}
\textbf{deleting user data} in an event-sourced system [...] \textbf{To which stream is the tombstone event written?}
The event stream of the specific user? Or is there an event stream specifically for tombstone events?
\end{displayquote}
\begin{displayquote}
\textbf{to keep some data} (e.g. the users' id), he advises to split the user stream into a \textbf{public and private event stream}. [...] How can you make sure that both events have been appended?
\end{displayquote}



Managing data privacy in the context of event streams accounts for 23.08\% of security challenges.
This suggests that developers find uncertainties on how to properly manage data privacy in events exchanged between microservices.

\noindent\textbf{Discussion:} We find that developers often manage data privacy in event streams using ad-hoc solutions, such as separating privacy and non-privacy data into different streams. However, these mechanisms do not enforce by design that data leakages do not occur, such as user data contained in events that are not supposed to be processed by certain consumer microservices. Besides, these mechanisms can lead to consistency problems when it is up to the developer to ensure privacy and non-privacy streams are in sync.

\vspace{1ex}
\begin{mdframed}
\textbf{Finding 8:} It is unclear for microservice developers how to properly safeguard security properties in the context of event processing in microservices.
\end{mdframed}
\vspace{1ex}

\subsubsection{Implications} 
Messaging technology providers could provide technology-specific guidelines on properly handling authentication and data privacy in their systems, accounting for the mechanisms that prevent exposing unintended event payloads. Cloud providers and framework developers can offer custom deployment templates to facilitate meeting certain security criteria. However, these should not be oblivious to the event management layer, suggesting that an appropriate cross-system integration is necessary to fulfill security challenges.

Microservice developers can, by principle, never include sensitive data in their generated events (i.e., privacy-by-design). However, this can jeopardize the decoupling benefit brought by event-based architecture, requiring consumers to contact the producer for the missing sensitive data. Researchers can extend existing code analysis tools to alert developers about bugs and configuration mistakes that can lead to unintended access to events and exposure of private data in event streams. Besides, researchers can investigate privacy-by-design event management methods without impact decoupling. For instance, protocols that automatically encrypt sensitive data in events without developer intervention.






\vspace{1ex}
\begin{mdframed}
\textbf{Summary:} Microservice developers encounter a myriad of problems when managing events, including dealing with large event payloads, evolving event schemas, auditing and validating security tokens through events, and processing events in order.
\end{mdframed}
\vspace{1ex}

\section{Threats to Validity}
\label{sec:validity}

In this section, we discuss the threats to the validity of our empirical study.


\noindent\textbf{Selection bias of the source.} 
Similar to previous research~\cite{wen2021empirical,bagherzadeh2019going,yang2016security}, our work uses SO as the data source to study the challenges developers encounter in practice. 
Although other data sources could supplement the findings, the heterogeneity of challenges, application scenarios, and developer expertise found in SO enables a trustworthy reflection of the state of the practice. For instance, the findings are aligned with several blog posts that report similar challenges~\cite{uber_microservice,uber_proxy,uber_payment, uber_money,nubank,nubank-arc,airbnb_double,airbnb_integrity,wix_pitfalls,wix_kafka}.



\noindent\textbf{Construction of tag set.} We select a set of tags to filter SO questions associated with event management in microservice architectures. The relevant keywords and metric applied may omit some tags associated with event management in microservices. To mitigate this threat, we rely on manual inspection conducted by independent researchers, and we adopt the lowest threshold used in previous work~\cite{wen2021empirical,bagherzadeh2019going,yang2016security} to include the highest amount of tags as possible.

\noindent\textbf{Subjective of researchers.} Since we adopt manual analysis to identify and categorize challenges, that may threaten our findings' validity. To minimize this threat, the first two authors analyzed and classified each filtered question in isolation. Upon conflicts, an experienced arbitrator coordinated the discussion to reach an agreement. 




\noindent\textbf{Generalizability of our findings.} We recognize that the application scenarios, requirements, design choices, event processing patterns, and the challenges extracted from SO may not be generalizable to all possible cloud and data platforms to which event-driven microservices can be deployed to. However, as SO is an industry-strength questions \& answers platform for developers, their questions tend to shed light on the most adopted industrial practices. Thus, the 10-year question range covered in this study strengthens our perception that the findings highlight a substantial portion of the challenges in the state of the practice.





\section{Related Work}
\label{subsec:related}

In this section, we summarize related work on challenges that developers face in microservice architectures.

\noindent\textbf{Benefits of microservice adoption.}
Zhang et al. ~\cite{zhang2019microservice,ZHOU2023111521} conduct interviews with microservice practitioners to understand the benefits brought about by microservice adoption. They find organizational transformation, decomposition, distributed monitoring, and bug localization as prominent challenges. Wang et al.~\cite{wang2021promises} interview and survey practitioners to collect and categorize best practices, challenges, and their related successful solutions employed by practitioners. They find managing API changes, lack of support for monitoring, and finding microservice granularity as common challenges.

\noindent\textbf{Microservice issues.}
Zhou et al.~\cite{zhou2021fault} performed a survey to characterize typical faults, debugging practices, and the challenges entailed by troubleshooting failures in microservice architectures. They find microservice developers can benefit from improved trace visualization tools, specially in cases related to microservice interactions. Ramírez et al.~\cite{ramirez2021empirical}
mine SO posts to identify common issues on
development and testing microservice applications. They find missing parameters as a prominent issue in communicating with other microservices, wrong library versions as the main impediment for service discovery, and connecting to other microservices with correct user credentials as an example of authentication \& authorization challenge. Waseem et al.~\cite{waseem2023understanding} employ a mixed-method empirical study to understand the types of issues microservice developers experience. Similarly to Ramírez et al.~\cite{ramirez2021empirical}, they find programming errors, missing artifacts, invalid configuration and communication as the main causes behind the issues. 

\noindent\textbf{Data Management in Microservices.}
Laigner et al.~\cite{vldb} characterize data management challenges in microservice architectures through a mixed-method empirical study. They focus on analyzing developers' pitfalls when implementing data management logic in the application layer and discussing the limitations of state-of-the-art database systems that prevent them from better serving microservice architectures.

\noindent\textbf{Event Sourcing Pattern.}
Overeem et al. ~\cite{OVEREEM2021110970} study the challenges practitioners experience using event sourcing. Through interviews with 25 engineers, they find event system evolution, steep learning curve, lack of available technology, rebuilding projections, and data privacy as the main issues affecting the development of systems that employ the event sourcing pattern. From 19 systems mentioned by interviewees, only 8 apply the microservice architectural style.

\noindent\textbf{Security in Microservices.}
Nasab et al.~\cite{nasab2023empirical} performs a mixed-method empirical study to understand the security practices in microservice systems. They collect 28 security practices open-source repositories and SO posts, which were later confirmed through a survey with practitioners.

\noindent\textbf{Exploratory Studies in Microservices.}
Hacaloglu and Demirors~\cite{event_software_size} study the usefulness of events for software size measurement. They find preliminary evidence that events can supplement software size measurement techniques. Lazzari and Farias~\cite{Farias_Lazzari_2023} reports an exploratory study that compares event-driven and REST architectural styles in the context of modularity. They find preliminary evidence that event-driven architecture improves the separation of concerns. 


Overall, we observe that challenges related to event management are overlooked in the literature. For instance, although few works report developers mentioning possible issues with message technology systems~\cite{waseem2023understanding,nasab2023empirical} and asynchronous task invocation ~\cite{ramirez2021empirical,zhou2021fault}, they are often addressed as general microservice configuration or invocation problems, preventing a proper characterization of the problem. Therefore, although event management in microservices has been rapidly gaining industry popularity~\cite{vldb,uber_microservice,uber_proxy,uber_payment,uber_money,nubank,nubank-arc,airbnb_double,airbnb_integrity,wix_pitfalls,wix_kafka}, it has not received due attention from the research community. In this work, we make the first attempt to investigate the specific challenges that developers face when managing events in microservice architectures.

\section{Conclusion}
\label{sec:conclusion}
    
In this paper, we mine and analyze several relevant Stack Overflow questions to characterize the state of the practice and the challenges microservice developers face while managing events. We identify key patterns developers use while attempting to realize their functional and non-functional requirements, suggesting a tension between achieving requirements related to data consistency, loose coupling, and performance. 

To further understand these tensions, we manually examined 628 sampled questions and identified key challenges microservice developers face in varied application scenarios. These include issues related to queuing, delivering, and processing events at the application level, as well as monitoring and securing the event flow. Based on the myriad of practical findings, we provide actionable implications for messaging systems, framework maintainers, cloud providers, and researchers. 

We hope that the results drive the reflection of microservice developers and researchers to escape from the dangers of asynchronous, event-based designs and to build event management technologies that meet the expectations of microservice developers respectively.

\begin{acks}
	We thank Jean Mello and Marcos Antonio Vaz Salles for early discussions and initial drafting of this manuscript.
	
	This work was supported by the PAPRICAS.org project - Independent Research Fund of Denmark (Number 9131-00077B). Part of the computation done for this project was performed on the UCloud interactive HPC system.
\end{acks}

\bibliographystyle{ACM-Reference-Format}
\bibliography{main}
  
\end{document}